\documentclass[english]{article}
\usepackage[T1]{fontenc}
\usepackage[latin9]{inputenc}
\usepackage{amssymb}
\usepackage{graphicx}
\usepackage{esint}

\makeatletter

\providecommand{\tabularnewline}{\\}

\makeatletter
 \@addtoreset{equation}{section}
 
\makeatother
\providecommand{\tabularnewline}{\\}

\newif\ifContLineOne
\newif\ifContLineTwo
\newif\ifContLineThree

\def\conC#1{\vbox{\ialign{##\crcr
  \ifContLineThree\hrulefill\else\vphantom{\hrulefill}\fi\crcr
  \noalign{\kern3.2pt\nointerlineskip}
  \ifContLineTwo\hrulefill\else\vphantom{\hrulefill}\fi\crcr
  \noalign{\kern3.2pt\nointerlineskip}
  \ifContLineOne\hrulefill\else\vphantom{\hrulefill}\fi\crcr
  \noalign{\nointerlineskip}
  $\hfil\textstyle{\vbox to 14pt{}#1}\hfil$\crcr}}}

\def\DrawLeg#1#2{
  \kern-.2pt              
  \dimen2 =#1             
  \advance\dimen2 by 2pt  
  \dimen3 = 10.6pt        
  \dimen4 =3.6pt          
  \advance\dimen3 by -\dimen2 
  \multiply\dimen4 by #2
  \advance\dimen3 by \dimen4
  \raise\dimen2 \hbox{\vrule height\dimen3 width .4pt} 
  \kern-.2pt}             

\def\begC#1#2{\setbox0 =\hbox{$\textstyle{#2}$}
  \dimen0=.5\wd0 \dimen1=\ht0
  \conC{\hskip\dimen0}
  \count255=#1
  \ifnum\count255 =1 \ContLineOnetrue\else
  \ifnum\count255 =2 \ContLineTwotrue\else
  \ifnum\count255 =3 \ContLineThreetrue\fi\fi\fi
  \DrawLeg{\dimen1}{\count255}
  \conC{\hskip\dimen0}
  \kern-\dimen0\kern-\dimen0 \box0}

\def\endC#1#2{\setbox0 =\hbox{$\textstyle{#2}$}
  \dimen0=.5\wd0 \dimen1=\ht0
  \conC{\hskip\dimen0}
  \count255=#1
  \ifnum\count255 =1 \ContLineOnefalse\else
  \ifnum\count255 =2 \ContLineTwofalse\else
  \ifnum\count255 =3 \ContLineThreefalse\fi\fi\fi
  \DrawLeg{\dimen1}{\count255}
  \conC{\hskip\dimen0}
  \kern-\dimen0\kern-\dimen0 \box0}

\makeatother

\usepackage{babel}
\begin{document}
\begin{titlepage}

\global\long\def\thefootnote{\fnsymbol{footnote}}
\begin{flushright}
\begin{tabular}{l}
UTHEP- 724\tabularnewline
\end{tabular}
\par\end{flushright}

\bigskip{}
\begin{center}
\textbf{\Large{}Multiloop Amplitudes of Light-cone Gauge String Field
Theory for Type II Superstrings}{\Large{} }
\par\end{center}{\Large \par}

\bigskip{}
\begin{center}
{\large{}{}Nobuyuki Ishibashi}\footnote{e-mail: ishibash@het.ph.tsukuba.ac.jp}
{\large{}{}}
\par\end{center}{\large \par}

\begin{center}
\textit{Tomonaga Center for the History of the Universe,}\\
\textit{ University of Tsukuba}\\
\textit{ Tsukuba, Ibaraki 305-8571, JAPAN}\\
 
\par\end{center}

\bigskip{}

\bigskip{}

\bigskip{}
\begin{abstract}
Feynman amplitudes of light-cone gauge string field theory for Type
II superstrings are shown to be equivalent to those of the covariant
first quantized formulation. In order to regularize the contact term
divergences, we consider the theory in a linear dilaton background
$\Phi_{\mathrm{dilaton}}=-iQX^{1}$. We show that the scattering amplitudes
are correctly reproduced in the limit $Q\to0$, even with Ramond sector
external lines.
\end{abstract}
\global\long\def\thefootnote{\arabic{footnote}}

\end{titlepage}

\pagebreak{}

\section{Introduction}

It has been a long standing problem to construct a string field theory
for closed superstrings, because of the problems with the method to
calculate multiloop amplitudes using the picture changing operators.
Recently, Sen has constructed a manifestly Lorentz invariant and BRST
invariant string field theory for closed superstrings \cite{Sen2016c,Sen2015b,Sen2015j,Sen2015d,Sen2018,Lacroix2017}
based on the recently established method \cite{Sen2015,Sen2015a}
to calculate multiloop amplitudes using the picture changing operators,
generalizing the string field theory \cite{Zwiebach1993,Zwiebach1993a}
for closed bosonic strings with a nonpolynomial action. 

Light-cone gauge closed string field theory is a string field theory
which involves only cubic interaction terms \cite{Mandelstam:1973jk,Mandelstam:1974hk,Kaku:1974zz,Kaku:1974xu,Cremmer:1974ej}.
The equivalence of the scattering amplitudes of the light-cone approach
and the covariant approach has been established for closed bosonic
strings \cite{Giddings:1986rf,D'Hoker:1987pr}. The light-cone gauge
amplitudes for superstrings suffer from the so-called contact term
divergences \cite{Greensite:1986gv,Greensite:1987sm,Greensite:1987hm}.
Up to these problems, the equivalence of the amplitudes for the two
approaches are shown in \cite{Aoki:1990yn} when all the external
lines are in the (NS,NS) sector for Type II superstrings or in the
NS sector for heterotic strings\footnote{In this paper, we deal with the light-cone gauge string field theory
in the NSR formalism. The light-cone gauge string field theory in
the Green-Schwarz formalism \cite{Green1983,Green1984d} can be shown
to be equivalent to that in the NSR formalism \cite{Mandelstam:1985wh}.}. 

In \cite{Baba:2009kr,Baba:2009ns,Baba:2009fi}, we show that the contact
term divergences are regularized by dimensional regularization. Light-cone
gauge string field theory can be formulated in noncritical dimensions
or taking the worldsheet theory for the transverse variables to be
the one with central charge $c\ne12$. One convenient choice of the
worldsheet theory is that in a linear dilaton background $\Phi_{\mathrm{dilaton}}=-iQX^{1}$,
with a space-like direction $X^{1}$ and a real constant $Q$ . Although
the theory becomes Lorentz noninvariant, it is equivalent to a BRST
invariant conformal gauge formulation with an unusual longitudinal
part. As in the conventional field theory, one can define the amplitudes
as analytic functions of $Q$ and take the limit $Q\to0$ to obtain
those in the critical dimensions. In \cite{Baba:2009zm,Ishibashi:2010nq,Ishibashi:2011fy,Ishibashi2011a},
we have found that the correct tree level amplitudes can be obtained
by taking the limit $Q\to0$ in the dimensionally regularized amplitudes
without adding any counterterms. In \cite{Ishibashi2017b,Ishibashi2018},
we have shown the same thing for multiloop amplitudes when all the
external lines are in the (NS,NS) or the NS sector. 

What we would like to do in this paper is to generalize the results
obtained so far to the case where some of the external lines are in
the (NS,R), (R,NS) or (R,R) sector, in the case of Type II superstrings.
Since our approach does not rely on the superspace formalism on the
worldsheet, it is straightforward to deal with the spin fields, which
are necessary for constructing amplitudes with Ramond sector external
lines. We show that the light-cone gauge amplitudes can be recast
into the conformal gauge ones in the same way as in \cite{Ishibashi2017b,Ishibashi2018}.
We find that the regularization works even in the presence of spin
fields and the correct amplitudes are obtained by taking the limit
$Q\to0$. The heterotic strings are discussed in a separate paper,
for the reasons explained in section \ref{sec:Conclusions-and-discussion}.

The organization of this paper is as follows. In section \ref{sec:Light-cone-gauge-amplitudes},
we present the form of the light-cone gauge amplitude as an integration
of a correlation function of vertex operators made from the transverse
variables, over the moduli parameters. In sections \ref{sec:Amplitudes-in-the},
\ref{sec:Correlation-functions-on} and \ref{sec:Equivalence-of-the},
we explain how we can transform the integrands of the light-cone gauge
amplitudes into correlation functions of conformal gauge worldsheet
theory and obtain the expression of the scattering amplitudes of the
covariant approach. In section \ref{sec:Dimensional-regularization},
we show that the dimensional regularization works as in the previous
papers and the desired results are obtained in the limit $Q\to0$.
Section \ref{sec:Conclusions-and-discussion} is devoted to discussions.
Several technical points are treated in the appendices. 

\section{Light-cone gauge amplitudes\label{sec:Light-cone-gauge-amplitudes}}

\begin{figure}
\begin{centering}
\includegraphics[scale=0.5]{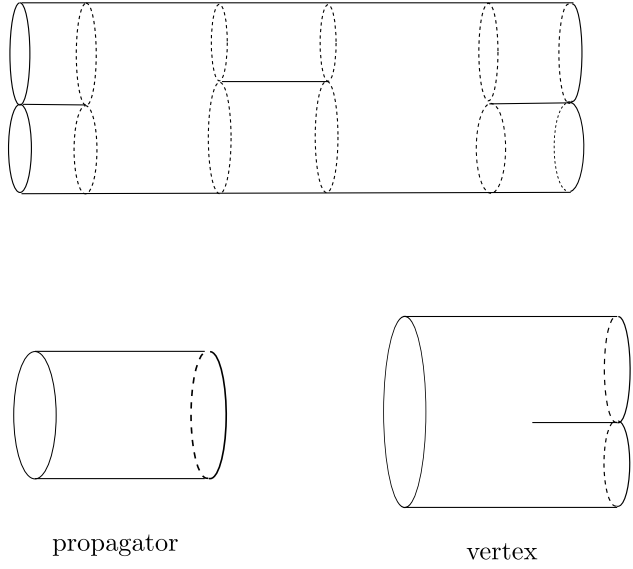}
\par\end{centering}
\caption{\label{fig:An-example-of}An example of light-cone diagram, propagator
and vertex.}
\end{figure}

A Feynman diagram in the light-cone gauge string field theory, as
schematically shown in figure \ref{fig:An-example-of}, is made from
the propagator and the vertex depicted in the same figure. A $g$-loop
$N$-point amplitude is given in the form
\begin{equation}
\mathcal{A}_{N}^{(g)}=(ig_{s})^{2g-2+N}C\int\prod_{K}dT_{K}\,\sum_{\mathrm{spin}\ \mathrm{structure}}F_{N}^{(g)}~,\label{eq:ANg}
\end{equation}
where $T_{K}\,\left(K=1,\cdots,6g-6+2N\right)$ denote the moduli
parameters of the light-cone diagram. In the case of Type II superstrings,
the integrand $F_{N}^{(g)}$ is given as a path integral over the
transverse variables on the light-cone diagram. In this paper, we
consider the flat background for simplicity and the transverse variables
are $X^{i}$, $\psi^{i}$, $\bar{\psi}^{i}$ $\left(i=1,\ldots,8\right)$.
The light-cone diagram can be regarded as a punctured Riemann surface
$\Sigma$. On $\Sigma$, there exists a unique holomorphic coordinate
$\rho$ whose real part is proportional to the light-cone time. $\rho$
can be expressed in terms of a local coordinate $z$ as \cite{Ishibashi:2013nma}
\begin{equation}
\rho(z)=\sum_{r=1}^{N}\alpha_{r}\left[\ln E(z,Z_{r})-2\pi i\int_{P_{0}}^{z}\omega\frac{1}{\mathop{\mathrm{Im}}\Omega}\mathop{\mathrm{Im}}\int_{P_{0}}^{Z_{r}}\omega\right]\,.\label{eq:rhoz}
\end{equation}
Here $E(z,w)$ is the prime form of the surface, $\omega$ is the
canonical basis of the holomorphic abelian differentials, $\Omega$
is the period matrix\footnote{For the mathematical background relevant for string perturbation theory,
we refer the reader to \cite{D'Hoker:1988ta}.} and $z=Z_{r}\,\left(r=1,\cdots,N\right)$ are the punctures. The
path integral on the light-cone diagram is defined by using the metric
\begin{equation}
ds^{2}=d\rho d\bar{\rho}\,.\label{eq:canonical-metric}
\end{equation}
This metric is not well-defined at the punctures, and the interaction
points of the light-cone diagram $z=z_{I}\,\left(I=2g-2+N\right)$,
which satisfy
\[
\partial\rho\left(z_{I}\right)=0\,.
\]

$F_{N}^{(g)}$ can be expressed in terms of correlation functions
defined with a metric $d\hat{s}^{2}=2\hat{g}_{z\bar{z}}dzd\bar{z}$
which is regular everywhere on the worldsheet, as 
\begin{eqnarray}
F_{N}^{(g)} & \propto & \left(2\pi\right)^{2}\delta^{2}\left(\sum p_{r}^{\pm}\right)e^{-\frac{1}{2}\Gamma\left[\varphi;\hat{g}_{z\bar{z}}\right]}\nonumber \\
 &  & \quad\times\int\left[dX^{i}d\psi^{i}d\bar{\psi}^{i}\right]_{\hat{g}_{z\bar{z}}}e^{-S^{\mathrm{LC}}}\prod_{I=1}^{2g-2+N}\left[\left|\partial^{2}\rho\left(z_{I}\right)\right|^{-\frac{3}{2}}T_{\mathrm{F}}^{\mathrm{LC}}\left(z_{I}\right)\bar{T}_{\mathrm{F}}^{\mathrm{LC}}\left(\bar{z}_{I}\right)\right]\prod_{r=1}^{N}V_{r}^{\mathrm{LC}}\left(Z_{r},\bar{Z}_{r}\right)\,.\nonumber \\
 &  & \ \label{eq:typeIIFN}
\end{eqnarray}
Here $S^{\mathrm{LC}}$ denotes the worldsheet action of the transverse
variables, $T_{\mathrm{F}}^{\mathrm{LC}},\,\bar{T}_{\mathrm{F}}^{\mathrm{LC}}$
are the supercurrents for these variables and $V_{r}^{\mathrm{LC}}$
denotes the vertex operator for the $r$-th external line inserted
at the puncture $z=Z_{r}$. The explicit form of the vertex operator
can be found in appendix \ref{subsec:Light-cone-gauge-vertex}. The
path integral measure $\left[dX^{i}d\psi^{i}d\bar{\psi}^{i}\right]_{\hat{g}_{z\bar{z}}}$
is defined with the metric $d\hat{s}^{2}=2\hat{g}_{z\bar{z}}dzd\bar{z}$.
Since the integrand is defined by using the metric (\ref{eq:canonical-metric}),
we need the Weyl anomaly factor $e^{-\frac{1}{2}\Gamma\left[\varphi;\hat{g}_{z\bar{z}}\right]}$,
where 
\begin{eqnarray}
 &  & \varphi=\ln\partial\rho\bar{\partial}\bar{\rho}-\ln\hat{g}_{z\bar{z}}\,,\nonumber \\
 &  & \Gamma\left[\varphi;\hat{g}_{z\bar{z}}\right]=-\frac{1}{4\pi}\int dz\wedge d\bar{z}\sqrt{\hat{g}}\left(\hat{g}^{ab}\partial_{a}\varphi\partial_{b}\varphi+2\hat{R}\varphi\right)\,.\label{eq:Gammasigma}
\end{eqnarray}
The right hand side of (\ref{eq:typeIIFN}) does not depend on the
choice of $\hat{g}_{z\bar{z}}$. The most convenient choice for $\hat{g}_{z\bar{z}}$
is the Arakelov metric $g_{z\bar{z}}^{\mathrm{A}}$ \cite{arakelov,Dugan:1987qe,D'Hoker:1989ae}.
The explicit form of $e^{-\Gamma\left[\varphi;g_{z\bar{z}}^{\mathrm{A}}\right]}$
is given as \cite{Ishibashi:2013nma}
\begin{eqnarray}
e^{-\Gamma\left[\varphi;g_{z\bar{z}}^{\mathrm{A}}\right]} & = & \prod_{r}\left[e^{-2\mathop{\mathrm{Re}}\bar{N}_{00}^{rr}}\left(2g_{Z_{r}\bar{Z}_{r}}^{\mathrm{A}}\right)^{-1}\right]\prod_{I}\left[\left|\partial^{2}\rho\left(z_{I}\right)\right|^{-3}\left(2g_{z_{I}\bar{z}_{I}}^{\mathrm{A}}\right)^{3}\right]\nonumber \\
 &  & \quad\times\exp\left[-2\sum_{I<J}G^{\mathrm{A}}\left(z_{I};z_{J}\right)-2\sum_{r<s}G^{\mathrm{A}}\left(Z_{r};Z_{s}\right)+2\sum_{I,r}G^{\mathrm{A}}\left(z_{I};Z_{r}\right)\right]\,.\label{eq:Gamma}
\end{eqnarray}
Here $G^{\mathrm{A}}\left(z;\,w\right)$ is the Arakelov Green's function
expressed as
\begin{equation}
G^{\mathrm{A}}\left(z;\,w\right)=-\ln\left|E(z,w)\right|^{2}+2\pi\mathrm{Im}\int_{w}^{z}\omega\frac{1}{\mathrm{Im}\Omega}\mathrm{Im}\int_{w}^{z}\omega-\frac{1}{2}\ln\left(2g_{z\bar{z}}^{\mathrm{A}}\right)-\frac{1}{2}\ln\left(2g_{w\bar{w}}^{\mathrm{A}}\right)\,,\label{eq:Arakelov}
\end{equation}
and
\[
\bar{N}_{00}^{rr}=\frac{\rho(z_{I^{(r)}})}{\alpha_{r}}-\sum_{s\neq r}\frac{\alpha_{s}}{\alpha_{r}}\ln E(Z_{r},Z_{s})+\frac{2\pi i}{\alpha_{r}}\int_{P_{0}}^{Z_{r}}\omega\frac{1}{\mathop{\mathrm{Im}}\Omega}\sum_{s=1}^{N}\alpha_{s}\mathop{\mathrm{Im}}\int_{P_{0}}^{Z_{s}}\omega~,
\]
where $z_{I^{\left(r\right)}}$ is defined to be the coordinate of
the interaction point at which the $r$-th external line interacts.

With the $e^{-\Gamma\left[\varphi;g_{z\bar{z}}^{\mathrm{A}}\right]}$
given by (\ref{eq:Gamma}), it is possible to write down the amplitude
(\ref{eq:typeIIFN}) explicitly using the formulas for correlation
functions of the transverse variables on higher genus Riemann surfaces
given in sections \ref{subsec:Correlation-functions-of-1} and \ref{subsec:Correlation-functions-of}. 

\section{Amplitudes in the conformal gauge\label{sec:Amplitudes-in-the}}

The amplitude (\ref{eq:ANg}) can be expressed by conformal gauge
worldsheet theory. Namely, as we will show in section \ref{sec:Equivalence-of-the},
(\ref{eq:typeIIFN}) is proportional to 
\begin{equation}
\int D\left[XBC\right]e^{-S^{\mathrm{tot}}}\prod_{K=1}^{6g-6+2N}\left[\oint_{C_{K}}\frac{dz}{\partial\rho}b_{zz}+\varepsilon_{K}\oint_{\bar{C}_{K}}\frac{d\bar{z}}{\bar{\partial}\bar{\rho}}b_{\bar{z}\bar{z}}\right]\prod_{I=1}^{2g-2+N}\left[X\left(z_{I}\right)\bar{X}\left(\bar{z}_{I}\right)\right]\prod_{r=1}^{N}V_{r}(Z_{r},\bar{Z}_{r})\,,\label{eq:confTypeII}
\end{equation}
with vertex operators $V_{r}(Z_{r},\bar{Z}_{r})$ whose explicit forms
will be given shortly. Here $S^{\mathrm{tot}}$ denotes the worldsheet
action for the conformal gauge variables, $D\left[XBC\right]$ denotes
the path integral measure for them, 
\begin{equation}
X\left(z\right)=\left[c\partial\xi-e^{\phi}T_{\mathrm{F}}+\frac{1}{4}\partial b\eta e^{2\phi}+\frac{1}{4}b\left(2\partial\eta e^{2\phi}+\eta\partial e^{2\phi}\right)\right](z)\label{eq:PCO}
\end{equation}
is the picture changing operator (PCO), $\bar{X}\left(\bar{z}\right)$
is its antiholomorphic counterpart and $T_{\mathrm{F}}$ denotes the
supercurrent for the matter part of the worldsheet theory. The contours
$C_{K}$ and $\varepsilon_{K}=\pm1$ are chosen so that the antighost
insertions correspond to the moduli parameters for the light-cone
diagram \cite{Ishibashi2017b}. The expression (\ref{eq:confTypeII})
coincides with the integrand of the amplitude obtained by the covariant
approach \cite{Verlinde1987b,Saroja1992}, in which the locations
of the PCO's are taken to be the interaction points of the light-cone
diagram\footnote{Notice that in the light-cone setup, the positions of the PCO's have
a fixed coordinate in the coordinate patch on the surface and we do
not need the $\partial\xi$ terms.} . 

The vertex operators $V_{r}(Z_{r},\bar{Z}_{r})$ are chosen from those
constructed in appendix \ref{subsec:Conformal-gauge-vertex}. Suppose
that the left moving part of $V_{r}^{\mathrm{LC}}$ for $r=1,\cdots,N-2M$
are in the NS sector and those for $r=N-2M+1,\cdots,N$ are in the
R sector. We use indices $s,s^{\prime},s^{\prime\prime}$ to denote
$r$ satisfying $1\leq r\leq N-2M,\,N-2M+1\leq r\leq N-M,\,N-M+1\leq r\leq N$
respectively. In constructing the conformal gauge amplitudes, we need
to treat the following two cases separately:
\begin{enumerate}
\item $M\ne0$ or the spin structure is even.
\item $M=0$ and the spin structure is odd.
\end{enumerate}
Although the case $M\ne0$ is our main concern in this paper, here
and in the following, we also discuss other cases for completeness.
If $M\ne0$ or the spin structure is even, the left moving part of
$V_{r}(Z_{r},\bar{Z}_{r})\,\left(r=1,\cdots N\right)$ denoted by
$V_{r\,\mathrm{L}}^{\left(p_{\mathrm{L}}\right)}(Z_{r})$ are taken
to be\footnote{The amplitude does not depend on which of the R sector vertex operators
are chosen to be in the $-\frac{1}{2}$ picture. }
\begin{eqnarray}
V_{s\,\mathrm{L}}^{\left(p_{\mathrm{L}}\right)}(Z_{s}) & = & V_{s\,\mathrm{L}}^{\left(-1\right)}(Z_{s})\,,\nonumber \\
V_{s^{\prime}\,\mathrm{L}}^{\left(p_{\mathrm{L}}\right)}(Z_{s^{\prime}}) & = & V_{s^{\prime}\,\mathrm{L}}^{\left(-\frac{1}{2}\right)}(Z_{s^{\prime}})\,,\nonumber \\
V_{s^{\prime\prime}\,\mathrm{L}}^{\left(p_{\mathrm{L}}\right)}(Z_{s^{\prime\prime}}) & = & V_{s^{\prime\prime}\,\mathrm{L}}^{\left(-\frac{3}{2}\right)}(Z_{s^{\prime\prime}})\,.\label{eq:VsLp}
\end{eqnarray}
When all the external lines are in the NS sector and the spin structure
is odd, two of the vertex operators should be taken to be in the pictures
$0$ and $-2$ \cite{Ishibashi2018}. For example, we take
\begin{eqnarray}
V_{1\,\mathrm{L}}^{\left(p_{\mathrm{L}}\right)}(Z_{1}) & = & V_{1\,\mathrm{L}}^{\left(-2\right)}(Z_{1})\,,\nonumber \\
V_{2\,\mathrm{L}}^{\left(p_{\mathrm{L}}\right)}(Z_{2}) & = & V_{2\,\mathrm{L}}^{\left(0\right)}(Z_{2})\,,\nonumber \\
V_{r\,\mathrm{L}}^{\left(p_{\mathrm{L}}\right)}(Z_{r}) & = & V_{r\,\mathrm{L}}^{\left(-1\right)}(Z_{r})\,\left(r\geq3\right)\,.\label{eq:VrLp}
\end{eqnarray}
The right moving part of the vertex operator $V_{r\,\mathrm{R}}^{\left(p_{\mathrm{R}}\right)}(Z_{r})\,\left(r=1,\cdots N\right)$
are defined following the same rule depending on $V_{r\,\mathrm{R}}^{\mathrm{LC}}$
and the spin structure of the right moving sector. Combining these,
we get the vertex operator $V_{r}(Z_{r},\bar{Z}_{r})$ expressed as
\[
V_{r}(Z_{r},\bar{Z}_{r})=V_{r\,\mathrm{L}}^{\left(p_{\mathrm{L}}\right)}(Z_{r})V_{r\,\mathrm{R}}^{\left(p_{\mathrm{R}}\right)}(\bar{Z}_{r})\,.
\]

\section{Correlation functions on higher genus Riemann surfaces\label{sec:Correlation-functions-on}}

In order to show that (\ref{eq:confTypeII}) is proportional to (\ref{eq:typeIIFN}),
we need to know the correlation functions of the longitudinal variables
$X^{\pm},\psi^{\pm}$ and the ghosts. These correlation functions
are derived from the so-called bosonization formula \cite{Verlinde:1986kw,AlvarezGaume:1987vm,Sonoda1987c}. 

The correlation functions for a free field theory with central charge
$c$ are recast into the form \cite{Belavin1986,Belavin1986a}
\begin{equation}
\left(\mbox{left moving part}\right)\times\left(\mbox{right moving part}\right)\times e^{-cS}\,,\label{eq:LR}
\end{equation}
using \cite{Ishibashi2017b}
\begin{equation}
\left(g_{z\bar{z}}^{A}\right)^{\frac{g}{2}}\exp\left[-\frac{2\pi}{g-1}\mathrm{Im}\int_{\left(g-1\right)z}^{\bigtriangleup}\omega\left(\mathrm{Im}\Omega\right)^{-1}\mathrm{Im}\int_{\left(g-1\right)z}^{\bigtriangleup}\omega\right]=\left|\sigma\left(z\right)\right|^{2}e^{\frac{3}{g-1}S}\,,\label{eq:gne1}
\end{equation}
for $g\ne1$ and
\begin{eqnarray}
\left(g_{z\bar{z}}^{A}\right)^{\frac{g}{2}}\exp\left[4\pi\mathrm{Im}\int_{P_{0}}^{z}\omega\left(\mathrm{Im}\Omega\right)^{-1}\mathrm{Im}\int_{P_{0}}^{\bigtriangleup}\omega\right] & \equiv & \left|\sigma\left(z\right)\right|^{2}e^{A}\,,\nonumber \\
\exp\left[-2\pi\mathrm{Im}\int_{P_{0}}^{\bigtriangleup}\omega\left(\mathrm{Im}\Omega\right)^{-1}\mathrm{Im}\int_{P_{0}}^{\bigtriangleup}\omega\right] & \equiv & e^{3S}\,,\label{eq:geq1}
\end{eqnarray}
for $g=1$. Here $\bigtriangleup$ is the Riemann class and $\sigma\left(z\right)$
is the holomorphic $\frac{g}{2}$ form which transforms as 
\begin{equation}
\sigma\left(z\right)\to e^{-2\pi i\int_{\left(g-1\right)z}^{\bigtriangleup}\omega_{J}+\pi i\left(g-1\right)\Omega_{JJ}}\sigma\left(z\right)\,,
\end{equation}
when $z$ is moved around the $B_{J}$ cycle once, and invariant when
$z$ is moved around the $A_{J}$ cycles. $S,\,A$ are quantities
which do not depend on $z$. Since the total central charge of the
worldsheet theory vanishes, we need only the left moving or right
moving part of the correlation function to calculate $F_{N}^{(g)}$.
In this section, we collect the necessary formulas to evaluate (\ref{eq:confTypeII}).

\subsection{Free bosons\label{subsec:Correlation-functions-of-1}}

Let us consider a free scalar boson $X$. The correlation functions
of local operators $e^{ipX}$ can be expressed as an integration of
the holomorphically factorized correlation function at the fixed internal
momenta \cite{D'Hoker:1988ta}:
\begin{eqnarray}
 &  & \int\left[dX\right]_{g_{z\bar{z}}^{\mathrm{A}}}e^{-S^{X}}\prod_{r}e^{ip_{r}X}\left(Z_{r},\bar{Z}_{r}\right)\nonumber \\
 &  & \quad\propto2\pi\delta\left(\sum p_{r}\right)e^{-S}\int\prod_{J}dP_{J}\left\langle \prod e^{ip_{r}X_{\mathrm{L}}}\left(Z_{r}\right)\right\rangle _{X_{\mathrm{L}},\,p_{J}}\left(\left\langle \prod e^{ip_{r}X_{\mathrm{L}}}\left(Z_{r}\right)\right\rangle _{X_{\mathrm{L}},\,p_{J}}\right)^{*}\,,\label{eq:scalarfact}
\end{eqnarray}
where $S^{X}$ denotes the action for $X$ and 
\begin{eqnarray*}
 &  & \left\langle \prod_{r}e^{ip_{r}X_{\mathrm{L}}}\left(Z_{r}\right)\right\rangle _{X_{\mathrm{L}},\,p_{J}}=\left\langle 1\right\rangle _{X_{\mathrm{L}}}\prod_{r>s}E\left(Z_{r},Z_{s}\right)^{p_{r}p_{s}}\exp\left[\pi i\sum_{J,J^{\prime}}P_{J}\Omega_{JJ^{\prime}}P_{J^{\prime}}+2\pi i\sum_{J}P_{J}\sum_{r}p_{r}\int_{P_{0}}^{Z_{r}}\omega_{J}\right]\,,\\
 &  & \left\langle 1\right\rangle _{X_{\mathrm{L}}}=\left(\frac{\prod_{i}E\left(z_{i},R\right)\sigma\left(R\right)\det\omega_{J}\left(z_{i}\right)}{\vartheta{0\atopwithdelims[]0}\left(\sum_{i}\int_{P_{0}}^{z_{i}}\omega-\int_{P_{0}}^{R}\omega-\int_{P_{0}}^{\bigtriangleup}\omega\right)\prod_{i>j}E\left(z_{i},z_{j}\right)\prod_{i}\sigma\left(z_{i}\right)}\right)^{\frac{1}{3}}\,.
\end{eqnarray*}
$X_{\mathrm{L}}$ here denotes the left moving part of $X$ and $\vartheta[\alpha](\zeta|\Omega)$
denotes the theta function with characteristics $[\alpha]={\alpha^{\prime}\atopwithdelims[]\alpha^{\prime\prime}}$.
Roughly speaking, $\left\langle 1\right\rangle _{X_{\mathrm{L}}}$
is the left moving part of the partition function of $X$ and it satisfies
\begin{eqnarray*}
\left(\frac{\det^{\prime}\left(-g^{\mathrm{A}\,z\bar{z}}\partial_{z}\partial_{\bar{z}}\right)}{\int d^{2}z\sqrt{g^{\mathrm{A}}}}\right)^{-\frac{1}{2}} & \propto & \int\prod_{J}dP_{J}\left|\left\langle 1\right\rangle _{X_{\mathrm{L}}}\exp\left[\pi i\sum P_{J}\Omega_{JJ^{\prime}}P_{J^{\prime}}\right]\right|^{2}e^{-S}\\
 & \propto & \left(\det\mathrm{Im}\Omega\right)^{-\frac{1}{2}}\left|\left\langle 1\right\rangle _{X_{\mathrm{L}}}\right|^{2}e^{-S}\,.
\end{eqnarray*}
Correlation functions involving derivatives of $X$ can be derived
from (\ref{eq:scalarfact}) in a holomorphically factorized form.

\subsection{Free fermions\label{subsec:Correlation-functions-of}}

Let us bosonize the free fermions $\psi^{\pm}$ so that
\[
\psi^{+}=e^{iH}\,,\psi^{-}=-e^{-iH}\,.
\]

For an even spin structure, the left moving part of the correlation
function of $\psi^{\pm}$ is given by \cite{Ishibashi2017b}
\begin{equation}
\left\langle \prod_{i=1}^{n}\psi^{-}(x_{i})\prod_{j=1}^{n}\psi^{+}(y_{j})\right\rangle _{\psi^{\pm}}=\left\langle 1\right\rangle _{X_{\mathrm{L}}}\vartheta[\alpha_{\mathrm{L}}]\left(0\right)\det\left[-S_{\alpha_{\mathrm{L}}}\left(x_{i},y_{j}\right)\right]\,,\label{eq:fermioncorr2}
\end{equation}
 where 
\begin{equation}
S_{\alpha}\left(z,w\right)=\frac{1}{E\left(z,w\right)}\frac{\vartheta\left[\alpha\right]\left(\int_{w}^{z}\omega\right)}{\vartheta\left[\alpha\right]\left(0\right)}\label{eq:Szego}
\end{equation}
is the Szeg\"{o} kernel and $\alpha_{\mathrm{L}}$ corresponds to
the spin structure of the fermion. When there are Ramond sector external
lines, the correlation function we need for calculating the amplitude
(\ref{eq:confTypeII}) is
\begin{eqnarray}
 &  & \left\langle \prod_{i=1}^{n}\psi^{-}(x_{i})\prod_{j=1}^{n}\psi^{+}(y_{j})\prod_{s^{\prime}}e^{-\frac{i}{2}H}\left(Z_{s^{\prime}}\right)\prod_{s^{\prime\prime}}e^{\frac{i}{2}H}\left(Z_{s^{\prime\prime}}\right)\right\rangle _{\psi^{\pm}}\nonumber \\
 &  & \quad=\left\langle 1\right\rangle _{X_{\mathrm{L}}}\vartheta[\alpha_{\mathrm{L}}]\left(\frac{1}{2}\sum_{s^{\prime}}\int_{P_{0}}^{Z_{s^{\prime}}}\omega-\frac{1}{2}\sum_{s^{\prime\prime}}\int_{P_{0}}^{Z_{s^{\prime\prime}}}\omega\right)\frac{\prod_{s^{\prime}<\tilde{s}^{\prime}}E\left(Z_{s^{\prime}},Z_{\tilde{s}^{\prime}}\right)^{\frac{1}{4}}\prod_{s^{\prime\prime}<\tilde{s}^{\prime\prime}}E\left(Z_{s^{\prime\prime}},Z_{\tilde{s}^{\prime\prime}}\right)^{\frac{1}{4}}}{\prod_{s^{\prime},s^{\prime\prime}}E\left(Z_{s^{\prime}},Z_{s^{\prime\prime}}\right)^{\frac{1}{4}}}\nonumber \\
 &  & \hphantom{\quad\sim\vartheta[\alpha_{\mathrm{L}}]\left(\frac{1}{2}\sum_{r^{\prime}=1}^{m}\int_{P_{0}}^{Z_{r^{\prime}}}\omega-\frac{1}{2}\sum_{r^{\prime\prime}=1}^{m}\int_{P_{0}}^{Z_{r^{\prime\prime}}}\omega\right)}\times\det\left(-S_{\alpha_{\mathrm{L}}}\left(x_{i},y_{j}\right)\right)\,,\label{eq:spincorr2}
\end{eqnarray}
where
\begin{eqnarray}
S_{\alpha_{\mathrm{L}}}\left(x,y\right) & = & \frac{\vartheta[\alpha_{\mathrm{L}}]\left(\int_{P_{0}}^{x}\omega-\int_{P_{0}}^{y}\omega+\frac{1}{2}\sum_{s^{\prime}}\int_{P_{0}}^{Z_{s^{\prime}}}\omega-\frac{1}{2}\sum_{s^{\prime\prime}}\int_{P_{0}}^{Z_{s^{\prime\prime}}}\omega\right)}{\vartheta[\alpha_{\mathrm{L}}]\left(\frac{1}{2}\sum_{s^{\prime}}\int_{P_{0}}^{Z_{s^{\prime}}}\omega-\frac{1}{2}\sum_{s^{\prime\prime}}\int_{P_{0}}^{Z_{s^{\prime\prime}}}\omega\right)}\nonumber \\
 &  & \quad\times\frac{1}{E\left(x,y\right)}\cdot\frac{\prod_{s^{\prime}}E\left(x,Z_{s^{\prime}}\right)^{\frac{1}{2}}\prod_{s^{\prime\prime}}E\left(y,Z_{s^{\prime\prime}}\right)^{\frac{1}{2}}}{\prod_{s^{\prime}}E\left(y,Z_{s^{\prime}}\right)^{\frac{1}{2}}\prod_{s^{\prime\prime}}E\left(x,Z_{s^{\prime\prime}}\right)^{\frac{1}{2}}}\,.\label{eq:spinprop}
\end{eqnarray}
$S_{\alpha_{\mathrm{L}}}\left(x,y\right)$ can be considered as the
propagator of the fermions in the presence of the spin fields. Notice
that $\vartheta[\alpha_{\mathrm{L}}]\left(\frac{1}{2}\sum_{s^{\prime}}\int_{P_{0}}^{Z_{s^{\prime}}}\omega-\frac{1}{2}\sum_{s^{\prime\prime}}\int_{P_{0}}^{Z_{s^{\prime\prime}}}\omega\right)\ne0$
for generic $Z_{s^{\prime}},Z_{s^{\prime\prime}}$ and there are no
problems about (\ref{eq:spincorr2}) even when $\alpha_{\mathrm{L}}$
corresponds to an odd spin structure.

When all the external lines are in the (NS,NS) sector and the spin
structure is odd, the formula we need is 
\begin{equation}
\left\langle \prod_{i=1}^{n}\psi^{-}(x_{i})\prod_{j=1}^{n}\psi^{+}(y_{j})\right\rangle _{\psi^{\pm}}=\left\langle 1\right\rangle _{X_{\mathrm{L}}}\int d\psi_{0}^{+}d\psi_{0}^{-}\det\left[-S_{\alpha_{\mathrm{L}}}\left(x_{i},y_{j}\right)\right]\,,\label{eq:freeodd}
\end{equation}
where 
\begin{equation}
S_{\alpha_{\mathrm{L}}}(x,y)=\psi_{0}^{-}h_{\alpha_{\mathrm{L}}}(x)\psi_{0}^{+}h_{\alpha_{\mathrm{L}}}(y)+\frac{1}{E\left(x,y\right)}\frac{\sum_{\nu}\partial_{\nu}\vartheta\left[\alpha_{\mathrm{L}}\right]\left(\int_{y_{j}}^{x_{i}}\omega\right)\omega_{\nu}(p)}{\sum_{\nu}\partial_{\nu}\vartheta\left[\alpha_{\mathrm{L}}\right]\left(0,\Omega\right)\omega_{\nu}(p)}\,,\label{eq:SalphaL}
\end{equation}
and
\[
h_{\alpha_{\mathrm{L}}}(z)=\sqrt{\sum_{J}\partial_{J}\vartheta\left[\alpha_{\mathrm{L}}\right]\left(0\right)\omega_{J}(z)}\,.
\]

\subsection{The reparametrization ghost}

The correlation function of the $bc$ system which appear in (\ref{eq:confTypeII})
is evaluated as 
\begin{eqnarray}
 &  & \int\left[dbd\bar{b}dcd\bar{c}\right]_{g_{z\bar{z}}^{\mathrm{A}}}e^{-S^{bc}}\prod_{r=1}^{N}\left[c(Z_{r})\bar{c}(\bar{Z}_{r})\right]\prod_{K=1}^{6g-6+2N}\left[\oint_{C_{K}}\frac{dz}{2\pi i}\frac{b}{\partial\rho}\left(z\right)+\varepsilon_{K}\oint_{\bar{C}_{K}}\frac{d\bar{z}}{2\pi i}\frac{\bar{b}}{\bar{\partial}\bar{\rho}}\left(\bar{z}\right)\right]\nonumber \\
 &  & \quad\propto\frac{\det^{\prime}\left(-g^{\mathrm{A}\,z\bar{z}}\partial_{z}\partial_{\bar{z}}\right)}{\int d^{2}z\sqrt{g^{\mathrm{A}}}}\prod_{r=1}^{N}\left(\alpha_{r}e^{2\mathop{\mathrm{Re}}\bar{N}_{00}^{rr}}\right)e^{-\Gamma\left[\varphi;g_{z\bar{z}}^{\mathrm{A}}\right]}\,.\label{eq:bc}
\end{eqnarray}
A proof of this formula was given in \cite{Ishibashi:2013nma}.

\subsection{The superghost}

The superghost system is bosonized so that
\[
\beta=e^{-\phi}\partial\xi,\,\gamma=\eta e^{\phi}\,.
\]
 The left moving part of the correlation function relevant to superstring
amplitudes involving Ramond sector external lines is 
\begin{eqnarray}
 &  & \left\langle \prod_{I}e^{\phi}\left(z_{I}\right)\prod_{s}e^{-\phi}\left(Z_{s}\right)\prod_{s^{\prime}}e^{-\frac{1}{2}\phi}\left(Z_{s^{\prime}}\right)\prod_{s^{\prime\prime}}e^{-\frac{3}{2}\phi}\left(Z_{s^{\prime\prime}}\right)\right\rangle _{\beta\gamma}\nonumber \\
 &  & \quad\sim\left[\left\langle 1\right\rangle _{X_{\mathrm{L}}}\vartheta[\alpha_{\mathrm{L}}]\left(\sum_{I}\int_{P_{0}}^{z_{I}}\omega-\sum_{r}\int_{P_{0}}^{Z_{s}}\omega-\frac{1}{2}\sum_{s^{\prime}}\int_{P_{0}}^{Z_{s^{\prime}}}\omega-\frac{3}{2}\sum_{s^{\prime\prime}}\int_{P_{0}}^{Z_{s^{\prime\prime}}}\omega-2\int_{P_{0}}^{\bigtriangleup}\omega\right)\right]^{-1}\nonumber \\
 &  & \hphantom{\quad\sim}\times\frac{\prod_{I,s}E\left(z_{I},Z_{s}\right)\prod_{I,s^{\prime}}E\left(z_{I},Z_{s^{\prime}}\right)^{\frac{1}{2}}\prod_{I,s^{\prime\prime}}E\left(z_{I},Z_{s^{\prime\prime}}\right)^{\frac{3}{2}}}{\prod_{I>J}E\left(z_{I},z_{J}\right)\prod_{s>\tilde{s}}E\left(Z_{s},Z_{\tilde{s}}\right)\prod_{s,s^{\prime}}E\left(Z_{s},Z_{s^{\prime}}\right)^{\frac{1}{2}}\prod_{s,s^{\prime\prime}}E\left(Z_{s},Z_{s^{\prime\prime}}\right)^{\frac{3}{2}}}\nonumber \\
 &  & \hphantom{\quad\sim}\times\frac{1}{\prod_{s^{\prime}>\tilde{s}^{\prime}}E\left(Z_{r^{\prime}},Z_{s^{\prime}}\right)^{\frac{1}{4}}\prod_{s^{\prime\prime}>\tilde{s}^{\prime\prime}}E\left(Z_{s^{\prime\prime}},Z_{\tilde{s}^{\prime\prime}}\right)^{\frac{9}{4}}\prod_{s^{\prime},s^{\prime\prime}}E\left(Z_{s^{\prime}},Z_{s^{\prime\prime}}\right)^{\frac{3}{4}}}\nonumber \\
 &  & \hphantom{\quad\sim}\times\frac{\prod_{s}\sigma\left(Z_{s}\right)^{2}\prod_{s^{\prime}}\sigma\left(Z_{s^{\prime}}\right)\prod_{s^{\prime\prime}}\sigma\left(Z_{s^{\prime\prime}}\right)^{3}}{\prod_{I}\sigma\left(z_{I}\right)^{2}}\,.\label{eq:betagammaRamond}
\end{eqnarray}
The formula relevant for odd spin structure amplitudes is 
\begin{eqnarray}
 &  & \left\langle \prod_{I}e^{\phi}\left(z_{I}\right)e^{-2\phi}\left(Z_{1}\right)\prod_{r\geq3}e^{-\phi}\left(Z_{r}\right)\right\rangle _{\beta\gamma}\nonumber \\
 &  & \quad\sim\left[\left\langle 1\right\rangle _{X_{\mathrm{L}}}\vartheta[\alpha_{\mathrm{L}}]\left(\sum_{I}\int_{P_{0}}^{z_{I}}\omega-2\int_{P_{0}}^{Z_{1}}\omega-\sum_{r\geq3}\int_{P_{0}}^{Z_{r}}\omega-2\int_{P_{0}}^{\bigtriangleup}\omega\right)\right]^{-1}\nonumber \\
 &  & \hphantom{\quad\sim}\times\frac{\prod_{I}E\left(z_{I},Z_{1}\right)^{2}\prod_{I,r\geq3}E\left(z_{I},Z_{r}\right)}{\prod_{I>J}E\left(z_{I},z_{J}\right)\prod_{r\geq3}E\left(Z_{r},Z_{1}\right)^{2}\prod_{r>s\geq3}E\left(Z_{r},Z_{s}\right)}\cdot\frac{\sigma\left(Z_{1}\right)^{4}\prod_{r\geq3}\sigma\left(Z_{r}\right)^{2}}{\prod_{I}\sigma\left(z_{I}\right)^{2}}\,.\label{eq:betagammaodd}
\end{eqnarray}

\subsection{Useful formulas}

Since $z_{I}\,\left(I=1,\cdots,2g-2+N\right)$ and $Z_{r}\,\left(r=1,\cdots,N\right)$
are the zeros and poles of the one form $\partial\rho\left(z\right)dz$
respectively, 

\begin{equation}
\sum_{I=1}^{2g-2+N}\int_{P_{0}}^{z_{I}}\vec{\omega}-\sum_{r=1}^{N}\int_{P_{0}}^{Z_{r}}\vec{\omega}-2\int_{P_{0}}^{\bigtriangleup}\vec{\omega}=\vec{m}+\Omega\vec{n}\,,\label{eq:divisor}
\end{equation}
with $\vec{m},\vec{n}\in\mathbb{Z}^{g}$. Using this, we get the following
expression of $e^{-\Gamma\left[\varphi;\hat{g}_{z\bar{z}}\right]}$
in (\ref{eq:Gamma})
\begin{equation}
e^{-\Gamma\left[\varphi;\hat{g}_{z\bar{z}}\right]}=\left|\left\langle 1\right\rangle _{\mathrm{LC}}\right|^{2}e^{24S}\,,\label{eq:Gammafactorized}
\end{equation}
where\footnote{Since $\rho$ and $z_{I}$ depend on the antiholomorphic moduli, it
may not be appropriate to identify $\left\langle 1\right\rangle _{\mathrm{LC}}$
with the left moving part of $e^{-\Gamma\left[\varphi;\hat{g}_{z\bar{z}}\right]}$.}
\begin{eqnarray}
\left\langle 1\right\rangle _{\mathrm{LC}} & = & \frac{\prod_{I>J}E\left(z_{I},z_{J}\right)^{2}\prod_{r>s}E\left(Z_{r},Z_{s}\right)^{2}}{\prod_{I,r}E\left(z_{I},Z_{r}\right)^{2}}\cdot\frac{\prod_{I}\sigma\left(z_{I}\right)^{4}}{\prod_{r}\sigma\left(Z_{r}\right)^{4}}\nonumber \\
 &  & \quad\times e^{-2\pi i\vec{n}\Omega\vec{n}}\prod_{r}e^{-\bar{N}_{00}^{rr}}\prod_{I}\partial^{2}\rho\left(z_{I}\right)^{-\frac{3}{2}}\,.\label{eq:1LC}
\end{eqnarray}

Other useful formulas can be derived from 
\[
\left|\partial\rho\left(z\right)\right|^{2}=C\,g_{z\bar{z}}^{\mathrm{A}}\exp\left[\sum_{r}G^{\mathrm{A}}\left(z;\,Z_{r}\right)-\sum_{I}G^{\mathrm{A}}\left(z;\,z_{I}\right)\right]\,,
\]
where $C$ is a quantity which does not depend on $z$ \cite{Ishibashi:2013nma}.
Substituting (\ref{eq:Arakelov}), (\ref{eq:gne1}), (\ref{eq:geq1})
and (\ref{eq:divisor}) into this, we get 
\begin{equation}
\partial\rho\left(z\right)=C^{\prime}\sigma\left(z\right)^{2}\frac{\prod_{I}E\left(z,z_{I}\right)}{\prod_{r}E\left(z,Z_{r}\right)}\exp\left[-2\pi i\vec{n}\cdot\int^{z}\vec{\omega}\right]\,,\label{eq:delrho}
\end{equation}
where $C^{\prime}$ is a quantity which does not depend on $z$. From
this formula, it is easy to get
\begin{eqnarray}
\alpha_{s} & = & \lim_{z\to Z_{s}}\left(z-Z_{s}\right)\partial\rho\left(z\right)\nonumber \\
 & = & C^{\prime}\sigma\left(Z_{s}\right)^{2}\frac{\prod_{I}E\left(Z_{s},z_{I}\right)}{\prod_{r\ne s}E\left(Z_{s},Z_{r}\right)}\exp\left[-2\pi i\vec{n}\cdot\int^{Z_{s}}\vec{\omega}\right]\,,\label{eq:alphas}\\
\partial^{2}\rho\left(z_{I}\right) & = & \lim_{z\to z_{I}}\frac{\partial\rho\left(z\right)}{z-z_{I}}\nonumber \\
 & = & C^{\prime}\sigma\left(z_{I}\right)^{2}\frac{\prod_{J\ne I}E\left(z_{I},z_{J}\right)}{\prod_{r}E\left(z_{I},Z_{r}\right)}\exp\left[-2\pi i\vec{n}\cdot\int^{z_{I}}\vec{\omega}\right]\,.\label{eq:d2rhozI}
\end{eqnarray}

\section{Equivalence of the light-cone gauge amplitude and the conformal gauge
amplitude\label{sec:Equivalence-of-the}}

With the formulas collected in the previous section, let us prove
that (\ref{eq:confTypeII}) is proportional to (\ref{eq:typeIIFN}).
It is possible to show that (\ref{eq:confTypeII}) is equal to 
\begin{equation}
\int D\left[XBC\right]e^{-S^{\mathrm{tot}}}\prod_{K=1}^{6g-6+2N}\left[\oint_{C_{K}}\frac{dz}{\partial\rho}b_{zz}+\varepsilon_{K}\oint_{\bar{C}_{K}}\frac{d\bar{z}}{\bar{\partial}\bar{\rho}}b_{\bar{z}\bar{z}}\right]\prod_{I=1}^{2g-2+N}\left[e^{\phi}T_{\mathrm{F}}^{\mathrm{LC}}\left(z_{I}\right)e^{\bar{\phi}}\bar{T}_{\mathrm{F}}^{\mathrm{LC}}\left(\bar{z}_{I}\right)\right]\prod_{r=1}^{N}V_{r}(Z_{r},\bar{Z}_{r})\,.\label{eq:conftypeII2}
\end{equation}
A proof is given in appendix \ref{subsec:A-proof-of}. Eq. (\ref{eq:confTypeII})
can be recast into the form
\begin{eqnarray}
 &  & \int\left[dX^{i}d\psi^{i}d\bar{\psi}^{i}\right]_{g_{z\bar{z}}^{\mathrm{A}}}e^{-S^{\mathrm{LC}}}\prod_{I}\left[T_{\mathrm{F}}^{\mathrm{LC}}\left(z_{I}\right)\bar{T}_{\mathrm{F}}^{\mathrm{LC}}\left(\bar{z}_{I}\right)\right]\nonumber \\
 &  & \hphantom{\int\left[dX^{i}\right]}\times\int D\left[X^{\pm}BC\right]_{g_{z\bar{z}}^{\mathrm{A}}}e^{-S^{X^{\pm}BC}}\prod_{I}\left[e^{\phi}\left(z_{I}\right)e^{\bar{\phi}}\left(\bar{z}_{I}\right)\right]\nonumber \\
 &  & \hphantom{\hphantom{\int\left[dX^{i}\right]}\times\int D\left[X^{\pm}BC\right]e^{-S^{X^{\pm}BC}}}\times\prod_{K}\left[\oint_{C_{K}}\frac{dz}{\partial\rho}b_{zz}+\varepsilon_{K}\oint_{\bar{C}_{K}}\frac{d\bar{z}}{\bar{\partial}\bar{\rho}}b_{\bar{z}\bar{z}}\right]\prod_{r}V_{r}(Z_{r},\bar{Z}_{r})\,.\label{eq:conftypeII3}
\end{eqnarray}
Here $S^{X^{\pm}BC}$ and $D\left[X^{\pm}BC\right]_{g_{z\bar{z}}^{\mathrm{A}}}$
denote the action and the path integral measure of the longitudinal
variables and the ghosts. 

From (\ref{eq:spincorr2}), (\ref{eq:freeodd}), (\ref{eq:betagammaRamond}),
(\ref{eq:betagammaodd}), (\ref{eq:1LC}), (\ref{eq:alphas}) and
(\ref{eq:d2rhozI}), we obtain formulas useful in performing the path
integral over the longitudinal variables and ghosts in (\ref{eq:conftypeII3}).
When there are vertex operators in the R sector or the spin structure
is even, we get
\begin{eqnarray}
 &  & \left\langle \prod_{s^{\prime}}e^{-\frac{i}{2}H}\left(Z_{s^{\prime}}\right)\prod_{s^{\prime\prime}}e^{\frac{i}{2}H}\left(Z_{s^{\prime\prime}}\right)\right\rangle _{\psi^{\pm}}\left\langle \prod_{I}e^{\phi}\left(z_{I}\right)\prod_{s}e^{-\phi}\left(Z_{s}\right)\prod_{s^{\prime}}e^{-\frac{1}{2}\phi}\left(Z_{s^{\prime}}\right)\prod_{s^{\prime\prime}}e^{-\frac{3}{2}\phi}\left(Z_{s^{\prime\prime}}\right)\right\rangle _{\beta\gamma}\nonumber \\
 &  & \quad=\pm\frac{\prod\alpha_{s^{\prime\prime}}^{\frac{1}{2}}}{\prod\alpha_{s^{\prime}}^{\frac{1}{2}}}\left(\left\langle 1\right\rangle _{\mathrm{LC}}\right)^{-\frac{1}{2}}\prod_{r}e^{-\frac{1}{2}\mathop{\mathrm{Re}}\bar{N}_{00}^{rr}}\prod_{I}\left(\partial^{2}\rho\left(z_{I}\right)\right)^{-\frac{3}{4}}\,,\label{eq:main}
\end{eqnarray}
and for odd spin structures we have
\begin{eqnarray}
 &  & \left\langle \psi^{+}\left(Z_{1}\right)\psi^{-}\left(Z_{2}\right)\right\rangle _{\psi^{\pm}}\left\langle \prod_{I}e^{\phi}\left(z_{I}\right)e^{-2\phi}\left(Z_{1}\right)\prod_{r\geq3}e^{-\phi}\left(Z_{r}\right)\right\rangle _{\beta\gamma}\nonumber \\
 &  & \quad=\pm\frac{\alpha_{1}}{\alpha_{2}}\left(\left\langle 1\right\rangle _{\mathrm{LC}}\right)^{-\frac{1}{2}}\prod_{r}e^{-\frac{1}{2}\mathop{\mathrm{Re}}\bar{N}_{00}^{rr}}\prod_{I}\left(\partial^{2}\rho\left(z_{I}\right)\right)^{-\frac{3}{4}}\,.\label{eq:odd}
\end{eqnarray}
(\ref{eq:bc}), (\ref{eq:main}) and (\ref{eq:odd}) imply 
\begin{eqnarray*}
 &  & \int D\left[X^{\pm}BC\right]_{g_{z\bar{z}}^{\mathrm{A}}}e^{-S^{X^{\pm}BC}}\prod\left[e^{\phi}\left(z_{I}\right)e^{\bar{\phi}}\left(\bar{z}_{I}\right)\right]\prod\left[\oint_{C_{K}}\frac{dz}{\partial\rho}b_{zz}+\varepsilon_{K}\oint_{\bar{C}_{K}}\frac{d\bar{z}}{\bar{\partial}\bar{\rho}}b_{\bar{z}\bar{z}}\right]\prod V_{r}(Z_{r},\bar{Z}_{r})\\
 &  & \quad\propto\left(2\pi\right)^{2}\delta^{2}\left(\sum p_{r}^{\pm}\right)e^{-\frac{1}{2}\Gamma\left[\varphi;\hat{g}_{z\bar{z}}\right]}\prod_{I}\left|\partial^{2}\rho\left(z_{I}\right)\right|^{-\frac{3}{2}}\prod_{r}V_{r}^{\mathrm{LC}}\left(Z_{r},\bar{Z}_{r}\right)\,.
\end{eqnarray*}
Substituting this into (\ref{eq:conftypeII3}), one can show that
the conformal gauge expression (\ref{eq:confTypeII}) is proportional
to the light-cone gauge expression (\ref{eq:typeIIFN}). Comparing
the factorization properties of these two expressions, one can fix
the proportionality constant. 

\section{Dimensional regularization\label{sec:Dimensional-regularization}}

Although the integrand $F_{N}^{\left(g\right)}$ in (\ref{eq:typeIIFN})
is proportional to the conformal gauge expression (\ref{eq:confTypeII}),
the amplitude (\ref{eq:ANg}) itself is divergent because of the contact
term divergences \cite{Greensite:1986gv,Greensite:1987sm,Greensite:1987hm}.
Fortunately, these are the only spurious singularities to worry about
in the light-cone gauge formulation. In the previous papers \cite{Ishibashi2017b,Ishibashi2018},
we have shown that it is possible to regularize the divergences by
dimensional regularization, if all the external lines are in the (NS,NS)
sector. In this section, we would like to show that the same results
hold for the other cases. 

The contact term divergences are regularized by taking the worldsheet
superconformal field theory to be the one with central charge $c\ne12$.
We take the worldsheet theory to be the one in a linear dilaton background
$\Phi_{\mathrm{dilaton}}=-iQX^{1}$, with a real constant $Q$ and
a space-like direction $X^{1}$ \cite{Ishibashi2017d}. The worldsheet
action of $X^{1}$ and its fermionic partners $\psi^{1},\bar{\psi}^{1}$
on a worldsheet with metric $ds^{2}=2\hat{g}_{z\bar{z}}dzd\bar{z}$
becomes 
\begin{eqnarray}
S\left[X^{1},\psi^{1},\bar{\psi}^{1};\hat{g}_{z\bar{z}}\right] & = & \frac{1}{8\pi}\int dz\wedge d\bar{z}\sqrt{\hat{g}}\left(\hat{g}^{ab}\partial_{a}X^{1}\partial_{b}X^{1}-2iQ\hat{R}X^{1}\right)\nonumber \\
 &  & \qquad+\frac{1}{4\pi}\int dz\wedge d\bar{z}i\left(\psi^{1}\bar{\partial}\psi^{1}+\bar{\psi}^{1}\partial\bar{\psi}^{1}\right)\,.\label{eq:linaction}
\end{eqnarray}
Since the number of fermionic variables $\psi^{i}$ and $\bar{\psi}^{i}$
does not depend on $Q$, we do not have difficulties in dealing with
chiral fermions in the regularization. 

It is straightforward to formulate the light-cone gauge string field
theory with such a worldsheet theory. The light-cone gauge amplitude
becomes (\ref{eq:ANg}) with 
\begin{eqnarray}
F_{N}^{(g)} & \propto & \left(2\pi\right)^{2}\delta^{2}\left(\sum p_{r}^{\pm}\right)e^{-\frac{1}{2}\left(1-Q^{2}\right)\Gamma\left[\varphi;\hat{g}_{z\bar{z}}\right]}\nonumber \\
 &  & \quad\times\int\left[dX^{i}d\psi^{i}d\bar{\psi}^{i}\right]_{\hat{g}_{z\bar{z}}}e^{-S^{\mathrm{LC}}}\prod_{I=1}^{2g-2+N}\left[\left|\partial^{2}\rho\left(z_{I}\right)\right|^{-\frac{3}{2}}T_{\mathrm{F}}^{\mathrm{LC}}\left(z_{I}\right)\bar{T}_{\mathrm{F}}^{\mathrm{LC}}\left(\bar{z}_{I}\right)\right]\prod_{r=1}^{N}V_{r}^{\mathrm{LC}}\left(Z_{r},\bar{Z}_{r}\right)\,.\nonumber \\
 &  & \ \label{eq:FNgLCtypeII}
\end{eqnarray}

In \cite{Ishibashi2017d}, it was shown that the amplitude given by
(\ref{eq:ANg}) with the integrand (\ref{eq:FNgLCtypeII}) is finite\footnote{In \cite{Ishibashi2017d}, it was shown that integrand becomes regular
at the possible singularities for $Q^{2}>10$. These singularities
may be harmless and the integral (\ref{eq:ANg}) may be well-defined
for $Q^{2}$ slightly less than $10$. } for $Q^{2}>10$. We use this light-cone gauge expression of the amplitude
for $Q^{2}>10$ to define it as an analytic function of $Q^{2}$,
which is denoted by $A^{\mathrm{LC}}\left(Q^{2}\right)$. The amplitude
in the critical dimensions may be given by the limit $\lim_{Q\to0}A^{\mathrm{LC}}\left(Q^{2}\right)$.
In order to study what happens in the limit $Q\to0$, we recast the
expression of the integrand into the one given by correlation functions
in the conformal gauge worldsheet theory. 

\subsection{Supersymmetric $X^{\pm}$ CFT}

As is explained in \cite{Baba:2009fi}, the light-cone gauge worldsheet
theory in the linear dilaton background corresponds to the conformal
gauge worldsheet theory with an unusual longitudinal part which is
called the supersymmetric $X^{\pm}$ CFT. The action of the supersymmetric
$X^{\pm}$ CFT for Type II superstrings is given in the form
\begin{equation}
S^{\mathcal{X}^{\pm}}=-\frac{1}{2\pi}\int d^{2}\mathbf{z}\left(\bar{D}\mathcal{X}^{+}D\mathcal{X}^{-}+\bar{D}\mathcal{X}^{-}D\mathcal{X}^{+}\right)-Q^{2}\Gamma_{\mathrm{super}}\left[\mathcal{X}^{+},\,\hat{g}_{z\bar{z}}\right]\,,\label{eq:SpmtypeII}
\end{equation}
where the supercoordinate $\mathbf{z}$ is given by 
\begin{equation}
\mathbf{z}=(z,\theta)\,,
\end{equation}
the superfield $\mathcal{X}^{\pm}$ is defined as 
\begin{equation}
\mathcal{X}^{\pm}\left(\mathbf{z},\bar{\mathbf{z}}\right)=X^{\pm}\left(z\right)+i\theta\psi^{\pm}\left(z\right)+i\bar{\theta}\bar{\psi}^{\pm}\left(\bar{z}\right)+\theta\bar{\theta}F^{\pm}\,,\label{eq:superfieldXpm}
\end{equation}
and 
\begin{eqnarray}
D & \equiv & \frac{\partial}{\partial\theta}+\theta\frac{\partial}{\partial z}\,,\nonumber \\
\bar{D} & \equiv & \frac{\partial}{\partial\bar{\theta}}+\bar{\theta}\frac{\partial}{\partial\bar{z}}\,,\nonumber \\
d^{2}\mathbf{z} & \equiv & d\left(\mathrm{Re}z\right)d\left(\mathrm{Im}z\right)d\theta d\bar{\theta}\,.
\end{eqnarray}
The interaction term $\Gamma_{\mathrm{super}}$ is given by 
\begin{eqnarray}
\Gamma_{\mathrm{super}}\left[\mathcal{X}^{+};\hat{g}_{z\bar{z}}\right] & = & -\frac{1}{2\pi}\int d^{2}\mathbf{z}\left(\bar{D}\Phi D\Phi+\theta\bar{\theta}\hat{g}_{z\bar{z}}\hat{R}\Phi\right)\,,\nonumber \\
\Phi\left(\mathbf{z},\bar{\mathbf{z}}\right) & = & \ln\left(\left(D\Theta^{+}\right)^{2}\left(\mathbf{z}\right)\left(\bar{D}\bar{\Theta}^{+}\right)^{2}\left(\bar{\mathbf{z}}\right)\right)-\ln\hat{g}_{z\bar{z}}\,,\label{eq:Phi}\\
\Theta^{+}\left(\mathbf{z}\right) & = & \frac{D\mathcal{X}^{+}}{(\partial\mathcal{X}^{+})^{\frac{1}{2}}}\left(\mathbf{z}\right)\,,\nonumber 
\end{eqnarray}
which is the super Liouville action defined for $\Phi$ with the background
metric $ds^{2}=2\hat{g}_{z\bar{z}}dzd\bar{z}$. 

When the spin structures are both even, the correlation functions
of the supersymmetric $X^{\pm}$ CFT are evaluated as \cite{Ishibashi2017b}
\begin{eqnarray}
 &  & \int\left[d\mathcal{X}^{+}d\mathcal{X}^{-}\right]_{\hat{g}_{z\bar{z}}}e^{-S^{\mathcal{X}^{\pm}}}\prod_{r}e^{-ip_{r}^{+}\mathcal{X}^{-}}(\mathbf{Z}_{r},\bar{\mathbf{Z}}_{r})\prod_{t}e^{-ip_{s}^{-}\mathcal{X}^{+}}(\mathbf{w}_{t},\bar{\mathbf{w}}_{t})\nonumber \\
 &  & \quad=\int\left[d\mathcal{X}^{+}d\mathcal{X}^{-}\right]_{\hat{g}_{z\bar{z}}}e^{-S_{\mathrm{\mathrm{free}}}^{\mathcal{X}^{\pm}}}\prod_{r}e^{-ip_{r}^{+}\mathcal{X}^{-}}(\mathbf{Z}_{r},\bar{\mathbf{Z}}_{r})\times e^{Q^{2}\Gamma_{\mathrm{super}}\left[\mathcal{X}^{+},\,\hat{g}_{z\bar{z}}\right]}\prod_{t}e^{-ip_{t}^{-}\mathcal{X}^{+}}(\mathbf{w}_{t},\bar{\mathbf{w}}_{t})\nonumber \\
 &  & \quad=(2\pi)^{2}\delta\left(\sum p_{t}^{-}\right)\delta\left(\sum p_{r}^{+}\right)\left(\frac{\det^{\prime}\left(-\hat{g}^{z\bar{z}}\partial_{z}\partial_{\bar{z}}\right)}{\int d^{2}z\sqrt{\hat{g}}}\right)^{-1}\left(\frac{\det^{\prime}\left(-\hat{g}^{z\bar{z}}\partial_{z}\partial_{\bar{z}}\right)}{\det\mathrm{Im}\Omega\int d^{2}z\sqrt{\hat{g}}}\right)^{-\frac{1}{2}}\vartheta[\alpha_{\mathrm{L}}]\left(0\right)\vartheta[\alpha_{\mathrm{R}}]\left(0\right)^{*}\nonumber \\
 &  & \hphantom{\quad=(2\pi)^{2}\delta\left(\sum_{s}p_{s}^{-}\right)\delta\left(\sum_{r}p_{r}^{+}\right)}\times\prod_{t}e^{-p_{t}^{-}\frac{\mathbf{\rho}_{\mathrm{s}}+\bar{\mathbf{\rho}}_{\mathrm{s}}}{2}}(\mathbf{w}_{t},\bar{\mathbf{w}}_{t})\,e^{Q^{2}\Gamma_{\mathrm{super}}\left[-\frac{i}{2}\left(\rho_{\mathrm{s}}+\bar{\rho}_{\mathrm{s}}\right),\ \hat{g}_{z\bar{z}}\right]}\,,\label{eq:superXpmCFTcorr}
\end{eqnarray}
where $S_{\mathrm{\mathrm{free}}}^{\mathcal{X}^{\pm}}$ is the free
action for the superfield $\mathcal{X}^{\pm}$. Regarding the second
line as a correlation function of
\[
e^{Q^{2}\Gamma_{\mathrm{super}}\left[\mathcal{X}^{+},\ \hat{g}_{z\bar{z}}\right]}\prod_{t}e^{-ip_{t}^{-}\mathcal{X}^{+}}(\mathbf{w}_{t},\bar{\mathbf{w}}_{t})
\]
for the free theory with the source term
\[
\prod_{r}e^{-ip_{r}^{+}\mathcal{X}^{-}}(\mathbf{Z}_{r},\bar{\mathbf{Z}}_{r})\,,
\]
we can calculate it by replacing the $\mathcal{X}^{+}\left(\mathbf{z},\bar{\mathbf{z}}\right)$
by its expectation value $-\frac{i}{2}\left(\rho_{\mathrm{s}}\left(\mathbf{z}\right)+\bar{\rho}_{\mathrm{s}}\left(\bar{\mathbf{z}}\right)\right)$
and derive the third line. Here $\rho_{s},\bar{\rho}_{s}$ are the
supersymmetric version of $\rho,\bar{\rho}$ and expressed as
\begin{eqnarray*}
\rho_{\mathrm{s}}\left(\mathbf{z}\right) & = & \rho\left(z\right)+\theta f\left(z\right)\,,\\
\bar{\rho}_{\mathrm{s}}(\bar{\mathbf{z}}) & = & \bar{\rho}\left(\bar{z}\right)+\bar{\theta}\bar{f}\left(\bar{z}\right)\,,
\end{eqnarray*}
with
\begin{eqnarray}
f\left(z\right) & = & -\sum_{r}\alpha_{r}\Theta_{r}S_{\alpha_{\mathrm{L}}}\left(z,Z_{r}\right)\,,\nonumber \\
\bar{f}\left(\bar{z}\right) & = & -\sum_{r}\alpha_{r}\bar{\Theta}_{r}S_{\alpha_{\mathrm{R}}}\left(\bar{z},\bar{Z}_{r}\right)\,.\label{eq:rhos}
\end{eqnarray}
$S_{\alpha_{\mathrm{L}}}$ and $S_{\alpha_{\mathrm{R}}}$ are taken
to be the Szeg\"{o} kernel (\ref{eq:Szego}). The explicit form of
$e^{-\Gamma_{\mathrm{super}}\left[-\frac{i}{2}\left(\rho_{\mathrm{s}}+\bar{\rho}_{\mathrm{s}}\right),\ \hat{g}_{z\bar{z}}^{\mathrm{A}}\right]}$
is given by
\begin{equation}
e^{-\Gamma_{\mathrm{super}}\left[-\frac{i}{2}\left(\rho_{\mathrm{s}}+\bar{\rho}_{\mathrm{s}}\right),\ \hat{g}_{z\bar{z}}^{\mathrm{A}}\right]}=\exp\left[-\frac{1}{2}\Gamma\left[\varphi;\hat{g}_{z\bar{z}}^{\mathrm{A}}\right]-\sum_{r}\left(\Delta\Gamma_{r\,\mathrm{L}}+\Delta\Gamma_{r\,\mathrm{R}}\right)-\sum_{I}\left(\Delta\Gamma_{I\,\mathrm{L}}+\Delta\Gamma_{I\,\mathrm{R}}\right)\right]\,,\label{eq:Gammasupermulti2}
\end{equation}
with
\begin{eqnarray*}
-\Delta\Gamma_{r\,\mathrm{L}} & = & \frac{1}{2\alpha_{r}}\frac{\partial ff}{\partial^{2}\rho}\left(z_{I^{\left(r\right)}}\right)\,,\\
-\Delta\Gamma_{r\,\mathrm{R}} & = & -\left(\Delta\Gamma_{r\,\mathrm{L}}\right)^{*}\,,\\
-\Delta\Gamma_{I\,\mathrm{L}} & = & \left\{ -\left(\frac{5}{12}\frac{\partial^{4}\rho}{\left(\partial^{2}\rho\right)^{3}}-\frac{3}{4}\frac{\left(\partial^{3}\rho\right)^{2}}{\left(\partial^{2}\rho\right)^{4}}\right)\partial ff+\frac{2}{3}\frac{\partial^{3}ff}{\left(\partial^{2}\rho\right)^{2}}-\frac{\partial^{3}\rho}{\left(\partial^{2}\rho\right)^{3}}\partial^{2}ff\right.\\
 &  & \left.\quad{}-\frac{1}{12}\frac{\partial^{3}f\partial^{2}f\partial ff}{\left(\partial^{2}\rho\right)^{4}}\right\} \left(z_{I}\right)\,,\\
-\Delta\Gamma_{I\,\mathrm{R}} & = & -\left(\Delta\Gamma_{I\,\mathrm{L}}\right)^{*}\,,
\end{eqnarray*}
and $\Gamma\left[\varphi;\hat{g}_{z\bar{z}}^{\mathrm{A}}\right]$
given in (\ref{eq:Gamma}). 

The correlation function (\ref{eq:superXpmCFTcorr}) can be expressed
as an integration of the holomorphically factorized correlation function
at the fixed internal momenta with respect to $X^{+}+\frac{i}{2}\left(\rho+\bar{\rho}\right),\,X^{-}$:
\begin{eqnarray}
 &  & (2\pi)^{2}\delta\left(\sum p_{t}^{-}\right)\delta\left(\sum p_{r}^{+}\right)e^{-\left(3+12Q^{2}\right)S}\nonumber \\
 &  & \quad\times\int\prod_{J}d^{2}P_{J}^{\pm}\left\langle \prod e^{-ip_{r}^{+}\mathcal{X}_{\mathrm{L}}^{-}}(\mathbf{Z}_{r})\prod e^{-ip_{s}^{-}\mathcal{X}_{\mathrm{L}}^{+}}(\mathbf{w}_{t})\right\rangle _{\mathcal{X}_{\mathrm{L}}^{\pm},\,p_{J}^{\pm}}\nonumber \\
 &  & \hphantom{\quad\times\int\prod_{J}d^{2}P_{J}^{\pm}}\times\left\langle \prod e^{-ip_{r}^{+}\mathcal{X}_{\mathrm{R}}^{-}}(\bar{\mathbf{Z}}_{r})\prod e^{-ip_{s}^{-}\mathcal{X}_{\mathrm{R}}^{+}}(\bar{\mathbf{w}}_{t})\right\rangle _{\mathcal{X}_{\mathrm{R}}^{\pm},\,p_{J}^{\pm}}\,,\label{eq:superXpmCFTfact}
\end{eqnarray}
where
\begin{eqnarray}
 &  & \left\langle \prod e^{-ip_{r}^{+}\mathcal{X}_{\mathrm{L}}^{-}}(\mathbf{Z}_{r})\prod e^{-ip_{s}^{-}\mathcal{X}_{\mathrm{L}}^{+}}(\mathbf{w}_{t})\right\rangle _{\mathcal{X}_{\mathrm{L}}^{\pm},\,p_{J}^{\pm}}\nonumber \\
 &  & \quad=\left(\left\langle 1\right\rangle _{\mathrm{LC}}\right)^{-\frac{Q^{2}}{2}}\left(\left\langle 1\right\rangle _{X_{\mathrm{L}}}\right)^{2}\exp\left[-2\pi i\sum_{J,J^{\prime}}P_{J}^{+}\Omega_{JJ^{\prime}}P_{J^{\prime}}^{-}\right]\left\langle 1\right\rangle _{\psi^{\pm}}\nonumber \\
 &  & \hphantom{\quad=}\times\exp\left[Q^{2}\left(\sum_{r}\Delta\Gamma_{r\,\mathrm{L}}+\sum_{I}\Delta\Gamma_{I\,\mathrm{L}}\right)-\frac{1}{2}\sum_{t}p_{t}^{-}\rho_{\mathrm{s}}\left(\mathbf{w}_{t}\right)\right]\,,\nonumber \\
 &  & \left\langle \prod e^{-ip_{r}^{+}\mathcal{X}_{\mathrm{R}}^{-}}(\bar{\mathbf{Z}}_{r})\prod e^{-ip_{s}^{-}\mathcal{X}_{\mathrm{R}}^{+}}(\bar{\mathbf{w}}_{t})\right\rangle _{\mathcal{X}_{\mathrm{R}}^{\pm},\,p_{J}^{\pm}}\nonumber \\
 &  & \quad=\left(\left\langle \prod e^{-ip_{r}^{+}\mathcal{X}_{\mathrm{L}}^{-}}(\mathbf{Z}_{r})\prod e^{-ip_{s}^{-}\mathcal{X}_{\mathrm{L}}^{+}}(\mathbf{w}_{t})\right\rangle _{\mathcal{X}_{\mathrm{L}}^{\pm},\,p_{J}^{\pm}}\right)^{*}\,.\label{eq:superXpmLR}
\end{eqnarray}

The correlation functions involving spin fields or those for odd spin
structure can be dealt with in the same way. With the spin fields,
we get
\begin{eqnarray}
 &  & \int\left[d\mathcal{X}^{+}d\mathcal{X}^{-}\right]_{\hat{g}_{z\bar{z}}}e^{-S_{\mathrm{super}}^{\pm}\left[\hat{g}_{z\bar{z}},\mathcal{X}^{\pm}\right]}\prod_{r}e^{-ip_{r}^{+}\mathcal{X}^{-}}(\mathbf{Z}_{r},\bar{\mathbf{Z}}_{r})\prod_{t}e^{-ip_{s}^{-}\mathcal{X}^{+}}(\mathbf{w}_{t},\bar{\mathbf{w}}_{t})\nonumber \\
 &  & \hphantom{\int\left[d\mathcal{X}^{+}d\mathcal{X}^{-}\right]_{\hat{g}_{z\bar{z}}}e^{-S_{\mathrm{super}}^{\pm}\left[\hat{g}_{z\bar{z}},\mathcal{X}^{\pm}\right]}}\times\prod_{s^{\prime}}e^{-\frac{i}{2}H}(Z_{s^{\prime}})\prod_{s^{\prime\prime}}e^{\frac{i}{2}H}(Z_{s^{\prime\prime}})\nonumber \\
 &  & \quad=\int\left[d\mathcal{X}^{+}d\mathcal{X}^{-}\right]_{\hat{g}_{z\bar{z}}}e^{-S_{\mathrm{\mathrm{free}}}\left[\hat{g}_{z\bar{z}},\mathcal{X}^{\pm}\right]}\prod_{r}e^{-ip_{r}^{+}\mathcal{X}^{-}}(\mathbf{Z}_{r},\bar{\mathbf{Z}}_{r})\prod_{t}e^{-ip_{t}^{-}\mathcal{X}^{+}}(\mathbf{w}_{t},\bar{\mathbf{w}}_{t})\nonumber \\
 &  & \hphantom{\quad=\int\left[d\mathcal{X}^{+}d\mathcal{X}^{-}\right]_{\hat{g}_{z\bar{z}}}e^{-S_{\mathrm{\mathrm{free}}}\left[\hat{g}_{z\bar{z}},\mathcal{X}^{\pm}\right]}}\times\prod_{s^{\prime}}e^{-\frac{i}{2}H}(Z_{s^{\prime}})\prod_{s^{\prime\prime}}e^{\frac{i}{2}H}(Z_{s^{\prime\prime}})\times e^{Q^{2}\Gamma_{\mathrm{super}}\left[-\frac{i}{2}\left(\rho_{\mathrm{s}}+\bar{\rho}_{\mathrm{s}}\right),\ \hat{g}_{z\bar{z}}^{\mathrm{A}}\right]}\,,\nonumber \\
\label{eq:XpmCFTfree2}
\end{eqnarray}
 where $S_{\alpha_{\mathrm{L}}}$ and $S_{\alpha_{\mathrm{R}}}$ in
(\ref{eq:rhos}) should be chosen from those given in \ref{subsec:Correlation-functions-of}
depending on the situation. It is straightforward to obtain the holomorphically
factorized correlation function at the fixed internal momenta
\begin{eqnarray}
 &  & \left\langle \prod e^{-ip_{s}^{+}\mathcal{X}_{\mathrm{L}}^{-}}(\mathbf{Z}_{s})\prod e^{-ip_{s^{\prime}}^{+}X_{\mathrm{L}}^{-}}e^{-\frac{i}{2}H}(Z_{s^{\prime}})\prod e^{-ip_{s^{\prime\prime}}^{+}X_{\mathrm{L}}^{-}}e^{\frac{i}{2}H}(Z_{s^{\prime\prime}})\prod e^{-ip_{t}^{-}\mathcal{X}_{\mathrm{L}}^{+}}(\mathbf{w}_{t})\right\rangle _{\mathcal{X}_{\mathrm{L}}^{\pm},\,p_{J}^{\pm}}\nonumber \\
 &  & \quad=\left(\left\langle 1\right\rangle _{\mathrm{LC}}\right)^{-\frac{Q^{2}}{2}}\left(\left\langle 1\right\rangle _{X_{\mathrm{L}}}\right)^{2}\exp\left[-2\pi i\sum_{J,J^{\prime}}P_{J}^{+}\Omega_{JJ^{\prime}}P_{J^{\prime}}^{-}\right]\nonumber \\
 &  & \hphantom{\quad=}\quad\times\left\langle \prod e^{-\frac{i}{2}H}(Z_{s^{\prime}})\prod e^{\frac{i}{2}H}(Z_{s^{\prime\prime}})\right\rangle _{\psi^{\pm}}\nonumber \\
 &  & \hphantom{\quad=}\quad\times\exp\left[Q^{2}\left(\sum_{r}\Delta\Gamma_{r\,\mathrm{L}}+\sum_{I}\Delta\Gamma_{I\,\mathrm{L}}\right)-\frac{1}{2}\sum_{t}p_{t}^{-}\rho_{\mathrm{s}}\left(\mathbf{w}_{t}\right)\right]\,.\label{eq:superXpmLR2}
\end{eqnarray}
When $\alpha_{\mathrm{L}}$ corresponds to an odd spin structure,
we get
\begin{eqnarray}
 &  & \left\langle \prod e^{-ip_{r}^{+}\mathcal{X}_{\mathrm{L}}^{-}}(\mathbf{Z}_{r})\prod e^{-ip_{s}^{-}\mathcal{X}_{\mathrm{L}}^{+}}(\mathbf{w}_{t})\right\rangle _{\mathcal{X}_{\mathrm{L}}^{\pm},\,p_{J}^{\pm}}\nonumber \\
 &  & \quad=\left(\left\langle 1\right\rangle _{\mathrm{LC}}\right)^{-\frac{Q^{2}}{2}}\left(\left\langle 1\right\rangle _{X_{\mathrm{L}}}\right)^{2}\exp\left[-2\pi i\sum_{J,J^{\prime}}P_{J}^{+}\Omega_{JJ^{\prime}}P_{J^{\prime}}^{-}\right]\nonumber \\
 &  & \hphantom{\quad=}\quad\times\int d\psi_{0}^{+}d\psi_{0}^{-}\exp\left[Q^{2}\left(\sum_{r}\Delta\Gamma_{r\,\mathrm{L}}+\sum_{I}\Delta\Gamma_{I\,\mathrm{L}}\right)-\frac{1}{2}\sum_{t}p_{t}^{-}\rho_{\mathrm{s}}\left(\mathbf{w}_{t}\right)\right]\,,\label{eq:superXpmLR1}
\end{eqnarray}
in which $S_{\alpha_{\mathrm{L}}}$ is taken to be the one given in
(\ref{eq:SalphaL}). The right moving part can be defined in the same
way. 

From these correlation functions, it is straightforward to check that
the theory for the longitudinal variables $X^{\pm},\psi^{\pm},\bar{\psi}^{\pm}$
is a superconformal field theory, in the same way as was done in \cite{Ishibashi2016a}.
The super stress tensor $T^{\mathcal{X}^{\pm}}(\mathbf{z})$ becomes
\begin{equation}
T^{\mathcal{X}^{\pm}}(\mathbf{z})=\frac{1}{2}\partial\mathcal{X}^{+}D\mathcal{X}^{-}+\frac{1}{2}\partial\mathcal{X}^{-}D\mathcal{X}^{+}+2Q^{2}S(\mathbf{z},\mbox{\mathversion{bold}\ensuremath{\mathcal{X}_{L}^{+}}})\,,\label{eq:TXpm}
\end{equation}
where $S(\mathbf{z},\mbox{\mathversion{bold}\ensuremath{\mathcal{X}_{L}^{+}}})$
denotes the super Schwarzian derivative 
\begin{equation}
S(\mathbf{z},\mbox{\mathversion{bold}\ensuremath{\mathcal{X}_{L}^{+}}})=\frac{\partial^{2}\Theta^{+}}{D\Theta}-2\frac{\partial D\Theta^{+}\partial\Theta^{+}}{\left(D\Theta^{+}\right)^{2}}~.
\end{equation}
$T^{\mathcal{X}^{\pm}}(\mathbf{z})$ satisfies the super Virasoro
algebra with the central charge 
\begin{equation}
c=3+12Q^{2}\,.\label{eq:central}
\end{equation}
With the ghost system and the stress tensor of the transverse variables,
one can construct the BRST charge which turns out to be nilpotent.
Therefore the regularization can be considered to be a gauge invariant
one.

\subsection{$Q\to0$}

The conformal gauge worldsheet theory corresponding to the light-cone
theory in noncritical dimensions consists of the supersymmetric $X^{\pm}$
CFT, the supersymmetric ghost system and the worldsheet theory of
the transverse variables. The integrand (\ref{eq:FNgLCtypeII}) can
be expressed in terms of correlation functions of the conformal gauge
worldsheet theory. It is possible to show that $F_{N}^{\left(g\right)}$
in (\ref{eq:FNgLCtypeII}) is proportional to 
\begin{eqnarray}
 &  & \int D\left[XBC\right]e^{-S^{\mathrm{tot}}}\prod_{K=1}^{6g-6+2N}\left[\oint_{C_{K}}\frac{dz}{\partial\rho}b_{zz}+\varepsilon_{K}\oint_{\bar{C}_{K}}\frac{d\bar{z}}{\bar{\partial}\bar{\rho}}b_{\bar{z}\bar{z}}\right]\prod_{I=1}^{2g-2+N}\left[X\left(z_{I}\right)\bar{X}\left(\bar{z}_{I}\right)\right]\nonumber \\
 &  & \hphantom{\int D\left[XBC\right]e^{-S^{\mathrm{tot}}}}\quad\times\prod_{r=1}^{N}\left[e^{-\frac{iQ^{2}}{\alpha_{r}}\mathcal{X}^{+}}\left(\hat{\tilde{\mathbf{z}}}_{I^{\left(r\right)}},\hat{\tilde{\bar{\mathbf{z}}}}_{I^{\left(r\right)}}\right)V_{r}(Z_{r},\bar{Z}_{r})\right]\,,\label{eq:FNgconftypeII}
\end{eqnarray}
where $\hat{\tilde{\mathbf{z}}}_{I^{\left(r\right)}},\,\hat{\tilde{\bar{\mathbf{z}}}}_{I^{\left(r\right)}}$
are the operator valued coordinates defined in \cite{Ishibashi2018}
and the vertex operators are those constructed in appendix \ref{subsec:Conformal-gauge-vertex}.
The proof goes in a way similar to the critical case. As is proved
in appendix \ref{subsec:A-proof-of}, the conformal gauge expression
(\ref{eq:FNgconftypeII}) is equal to 
\begin{eqnarray}
 &  & \int D\left[XBC\right]e^{-S^{\mathrm{tot}}}\prod_{K=1}^{6g-6+2N}\left[\oint_{C_{K}}\frac{dz}{\partial\rho}b_{zz}+\varepsilon_{K}\oint_{\bar{C}_{K}}\frac{d\bar{z}}{\bar{\partial}\bar{\rho}}b_{\bar{z}\bar{z}}\right]\prod_{I=1}^{2g-2+N}\left[e^{\phi}T_{\mathrm{F}}^{\mathrm{LC}}\left(z_{I}\right)e^{\bar{\phi}}\bar{T}_{\mathrm{F}}^{\mathrm{LC}}\left(\bar{z}_{I}\right)\right]\nonumber \\
 &  & \hphantom{\int D\left[XBC\lambda\right]e^{-S^{\mathrm{tot}}}}\quad\times\prod_{r=1}^{N}\left[e^{-\frac{iQ^{2}}{\alpha_{r}}\mathcal{X}^{+}}\left(\hat{\tilde{\mathbf{z}}}_{I^{\left(r\right)}},\hat{\tilde{\bar{\mathbf{z}}}}_{I^{\left(r\right)}}\right)V_{r}(Z_{r},\bar{Z}_{r})\right]\nonumber \\
 &  & \quad\propto\int\left[dX^{i}d\psi^{i}d\bar{\psi}^{i}\right]e^{-S^{\mathrm{LC}}}\prod_{I}\left[T_{\mathrm{F}}^{\mathrm{LC}}\left(z_{I}\right)\bar{T}_{\mathrm{F}}^{\mathrm{LC}}\left(\bar{z}_{I}\right)\right]\nonumber \\
 &  & \hphantom{\quad=\int\left[dX^{i}\right]}\times\int D\left[X^{\pm}BC\right]e^{-S^{X^{\pm}BC}}\prod_{K}\left[\oint_{C_{K}}\frac{dz}{\partial\rho}b_{zz}+\varepsilon_{K}\oint_{\bar{C}_{K}}\frac{d\bar{z}}{\bar{\partial}\bar{\rho}}b_{\bar{z}\bar{z}}\right]\prod_{I}\left[e^{\phi}\left(z_{I}\right)e^{\bar{\phi}}\left(\bar{z}_{I}\right)\right]\nonumber \\
 &  & \hphantom{\quad=\hphantom{\int\left[dX^{i}\right]}\times\int D\left[X^{\pm}BC\right]e^{-S^{X^{\pm}BC}}}\times\prod_{r=1}^{N}\left[e^{-\frac{iQ^{2}}{\alpha_{r}}\mathcal{X}^{+}}\left(\hat{\tilde{\mathbf{z}}}_{I^{\left(r\right)}},\hat{\tilde{\bar{\mathbf{z}}}}_{I^{\left(r\right)}}\right)V_{r}(Z_{r},\bar{Z}_{r})\right]\,.\label{eq:FNgconftypeII2}
\end{eqnarray}
Using (\ref{eq:bc}), (\ref{eq:betagammaRamond}), (\ref{eq:betagammaodd}),
(\ref{eq:superXpmLR1}) and (\ref{eq:superXpmLR2}), we obtain
\begin{eqnarray*}
 &  & \int D\left[X^{\pm}BC\right]e^{-S^{X^{\pm}BC}}\prod\left[\oint_{C_{K}}\frac{dz}{\partial\rho}b_{zz}+\varepsilon_{K}\oint_{\bar{C}_{K}}\frac{d\bar{z}}{\bar{\partial}\bar{\rho}}b_{\bar{z}\bar{z}}\right]\prod\left[e^{\phi}\left(z_{I}\right)e^{\bar{\phi}}\left(\bar{z}_{I}\right)\right]\\
 &  & \hphantom{\int D\left[X^{\pm}BC\right]e^{-S^{X^{\pm}BC}}}\times\prod_{r=1}^{N}\left[e^{-\frac{iQ^{2}}{\alpha_{r}}\mathcal{X}^{+}}\left(\hat{\tilde{\mathbf{z}}}_{I^{\left(r\right)}},\hat{\tilde{\bar{\mathbf{z}}}}_{I^{\left(r\right)}}\right)V_{r}(Z_{r},\bar{Z}_{r})\right]\\
 &  & \quad\propto\left(2\pi\right)^{2}\delta^{2}\left(\sum p_{r}^{\pm}\right)\\
 &  & \quad\hphantom{\quad\propto}\times\sum_{\mathrm{spin}\ \mathrm{structure}}e^{-\frac{1}{2}\left(1-Q^{2}\right)\Gamma\left[\varphi;\hat{g}_{z\bar{z}}\right]}\prod\left|\partial^{2}\rho\left(z_{I}\right)\right|^{-\frac{3}{2}}\prod V_{r}^{\mathrm{LC}}\left(Z_{r},\bar{Z}_{r}\right)\,.
\end{eqnarray*}
Substituting this into (\ref{eq:FNgconftypeII2}), one can see that
(\ref{eq:FNgconftypeII}) is proportional to (\ref{eq:FNgLCtypeII}). 

With the conformal gauge expression (\ref{eq:FNgconftypeII}), we
can show that the limit $\lim_{Q\to0}A^{\mathrm{LC}}\left(Q^{2}\right)$
coincides with the result of the covariant approach in the same way
as was done in previous papers \cite{Ishibashi2017b,Ishibashi2018}.
In the covariant formulation, the amplitudes can be obtained by the
method given by Sen and Witten in \cite{Sen2015,Sen2015a} using the
PCO's. Even with $Q\ne0$, since the vertex operators, PCO's and the
insertions $e^{-\frac{iQ^{2}}{\alpha_{r}}\mathcal{X}^{+}}\left(\hat{\tilde{\mathbf{z}}}_{I^{\left(r\right)}},\hat{\tilde{\bar{\mathbf{z}}}}_{I^{\left(r\right)}}\right)$
commute with the BRST charge, the conformal gauge expression (\ref{eq:FNgconftypeII})
can be deformed to define the amplitudes following the Sen-Witten
prescription. We can divide the moduli space into patches and put
the PCO's avoiding the spurious singularities as was explained in
\cite{Sen2015} and define the amplitude $A^{\mathrm{SW}}\left(Q^{2}\right)$.
Moving the locations of the PCO's, the amplitudes change by total
derivative terms in moduli space. For $Q^{2}$ large enough, these
total derivative terms do not contribute to the amplitudes, because
the infrared divergences are regularized. Therefore 
\[
A^{\mathrm{SW}}\left(Q^{2}\right)=A^{\mathrm{LC}}\left(Q^{2}\right)\,,
\]
as an analytic function of $Q^{2}$. Since $A^{\mathrm{SW}}\left(Q^{2}\right)$
is free from the spurious singularities, it can be well-defined for
$Q^{2}<10$ and 
\[
\lim_{Q\to0}A^{\mathrm{LC}}\left(Q^{2}\right)=A^{\mathrm{SW}}\left(0\right)\,,
\]
if the right hand side is well-defined. $A^{\mathrm{SW}}\left(0\right)$
is exactly the amplitude in the critical dimensions obtained by the
covariant approach.

\section{Conclusions\label{sec:Conclusions-and-discussion}}

In this paper, we have shown that the Feynman amplitudes of the light-cone
gauge string field theory for Type II superstrings coincide with those
of the covariant first quantized approach, even with Ramond sector
external lines. The divergences due to the collisions of the supercurrents
inserted at the interaction points are regularized by taking the worldsheet
theory of the transverse variables to be the one with a linear dilaton
background $\Phi_{\mathrm{dilaton}}=-iQX^{1}$. The amplitudes are
defined as analytic functions of $Q^{2}$ and in the limit $Q\to0$
they coincide with those of the critical string calculated by using
the Sen-Witten prescription. 

With the formulas given in this paper, it should be possible to do
the same thing for heterotic strings and Type I strings. A problem
with heterotic strings is the holomorphic factorization. The integrand
$F_{N}^{\left(g\right)}$ for heterotic strings should be given as
a product of the holomorphic part of that for Type II superstrings
and the antiholomorphic part of that for bosonic strings. Such a holomorphic
factorization is subtle in the formulation of light-cone gauge string
field theory, because the $\rho$ coordinate depends on the antiholomorphic
moduli parameters. We will discuss this issue in a separate publication. 

We have shown that the scattering amplitudes of Type II strings can
be reproduced by the light-cone gauge string field theory with only
cubic interaction terms. The regularization we propose regularizes
the infrared divergences of superstring theory in a gauge invariant
way. With such a formulation of superstring theory, we should be able
to study nonperturbative dynamics of the theory. We will leave this
subject for future investigation. 

\section*{Acknowledgments}

We wish to thank Koichi Murakami for discussions. We are also grateful
to the organizers of Nishinomiya-Yukawa memorial workshop \textquotedbl{}New
Frontiers in String Theory 2018\textquotedbl{} for hospitality. This
work was supported in part by Grant-in-Aid for Scientific Research
(C) (JP25400242) and (18K03637) from MEXT.

\appendix

\section{Vertex operators}

In this appendix, we present the forms of the vertex operators which
are used in the main text. Here we consider the vertex operators in
the noncritical case $Q\ne0$. Those in the critical case can be obtained
by putting $Q=0$. 

\subsection{Light-cone gauge vertex operators\label{subsec:Light-cone-gauge-vertex}}

The light-cone vertex operators are local fields corresponding to
the external states. The vertex operator $V_{r}^{\mathrm{LC}}\left(Z_{r},\bar{Z}_{r}\right)$
corresponding to the $r$-th external state $\left|r\right\rangle $
given by
\[
\left|r\right\rangle =\left|r\right\rangle _{\mathrm{L}}\otimes\left|r\right\rangle _{\mathrm{R}}\,,
\]
as a tensor product of left and right moving states, can be given
in the factorized form
\begin{equation}
V_{r}^{\mathrm{LC}}\left(Z_{r},\bar{Z}_{r}\right)=V_{r\,\mathrm{L}}^{\mathrm{LC}}\left(Z_{r}\right)V_{r\,\mathrm{R}}^{\mathrm{LC}}\left(\bar{Z}_{r}\right)\,,\label{eq:VrLC}
\end{equation}
accordingly. 

For a left moving state 
\[
\alpha_{-n_{1}}^{i_{1}(r)}\cdots\psi_{-s_{1}}^{j_{1}(r)}\cdots\left|\vec{p}_{r}\right\rangle 
\]
in the NS sector, the vertex operator is given by
\begin{eqnarray}
V_{r\,\mathrm{L}}^{\mathrm{LC}}\left(Z_{r}\right) & = & \left(\alpha_{r}\right)^{\frac{1}{2}}\oint_{Z_{r}}\frac{dz}{2\pi i}i\partial\tilde{X}^{i_{1}}\left(z\right)w_{r}^{-n_{1}}\cdots\oint_{Z_{r}}\frac{dz}{2\pi i}\left(\frac{\partial w_{r}}{\partial z}\right)^{\frac{1}{2}}\psi^{j_{1}}\left(z\right)w_{r}^{-s_{1}-\frac{1}{2}}\cdots\nonumber \\
 &  & \hphantom{\alpha_{r}^{\frac{1}{2}}\oint_{Z_{r}}\frac{dz}{2\pi i}i\partial X^{i_{1}}\left(z\right)w_{r}^{-n_{1}}\cdots}\times e^{i\vec{p}_{r}\cdot\vec{\tilde{X}}_{\mathrm{L}}}\left(Z_{r}\right)\left(\frac{\partial w_{r}}{\partial z}\right)^{-\frac{1}{2}\left|\vec{p}_{r}\right|^{2}}e^{-\frac{1}{2}p_{r}^{-}\rho\left(z_{I^{\mathrm{\left(r\right)}}}\right)}\,,\label{eq:LCNS}
\end{eqnarray}
where 
\begin{eqnarray*}
\tilde{X}^{i} & \equiv & X^{i}-iQ\delta^{i1}\ln(2g_{z\bar{z}})\,,\\
w_{r} & \equiv & \exp\left[\frac{1}{\alpha_{r}}\left(\rho\left(z\right)-\rho(z_{I^{(r)}})\right)\right]\,.
\end{eqnarray*}
 $\tilde{X}_{\mathrm{L}}^{i}\left(z\right)$ denotes the left moving
part of the variable $\tilde{X}^{i}\left(z,\bar{z}\right)$. The momentum
$p_{r}$ satisfies the on-shell condition 
\[
\frac{1}{2}\left(-2p_{r}^{+}p_{r}^{-}+p_{r}^{i}p_{r}^{i}\right)+Qp_{r}^{1}+\mathcal{N}_{r}=\frac{1}{2}\left(1-Q^{2}\right)~,\quad\mathcal{N}_{r}\equiv\sum_{k}n_{k}+\sum_{l}s_{l}~,
\]
with $n_{k}\in\mathbb{Z},\,s_{l}\in\mathbb{Z}+\frac{1}{2}$. The factor
$\left(\alpha_{r}\right)^{\frac{1}{2}}$ on the right hand side of
(\ref{eq:LCNS}) comes from the normalization of the three string
vertex and is essential for the Lorentz invariance in the critical
case. 

For a left moving state 
\[
\alpha_{-n_{1}}^{i_{1}(r)}\cdots\psi_{-s_{1}}^{j_{1}(r)}\cdots\left|\vec{p}_{r},\alpha\right\rangle 
\]
in the R sector,
\begin{eqnarray}
V_{r\,\mathrm{L}}^{\mathrm{LC}}\left(Z_{r}\right) & = & \alpha_{r}\oint_{Z_{r}}\frac{dz}{2\pi i}i\partial\tilde{X}^{i_{1}}\left(z\right)w_{r}^{-n_{1}}\cdots\oint_{Z_{r}}\frac{dz}{2\pi i}\left(\frac{\partial w_{r}}{\partial z}\right)^{\frac{1}{2}}\psi^{j_{1}}\left(z\right)w_{r}^{-s_{1}-\frac{1}{2}}\cdots\nonumber \\
 &  & \hphantom{\left(\alpha_{r}\right)^{\frac{1}{2}}}\times S_{\alpha}e^{i\vec{p}_{r}\cdot\vec{\tilde{X}}_{\mathrm{L}}}\left(Z_{r}\right)\left(\frac{\partial w_{r}}{\partial z}\right)^{-\frac{1}{2}\left|\vec{p}_{r}\right|^{2}-\frac{1}{2}}e^{-\frac{1}{2}p_{r}^{-}\rho\left(z_{I^{\mathrm{\left(r\right)}}}\right)}\,,\label{eq:LCR}
\end{eqnarray}
where $S_{\alpha}$ is the spin field and the on-shell condition is
\[
\frac{1}{2}\left(-2p_{r}^{+}p_{r}^{-}+p_{r}^{i}p_{r}^{i}\right)+Qp_{r}^{1}+\mathcal{N}_{r}=-\frac{Q^{2}}{2}~,\quad\mathcal{N}_{r}\equiv\sum_{k}n_{k}+\sum_{l}m_{l}~,
\]
with $n_{k}\in\mathbb{Z},\,s_{l}\in\mathbb{Z}$. 

The right moving part of the vertex operator $V_{r\,\mathrm{R}}^{\mathrm{LC}}\left(\bar{Z}_{r}\right)$
can be defined in the same way. 

\subsection{Conformal gauge vertex operators\label{subsec:Conformal-gauge-vertex}}

We define the conformal gauge vertex operators corresponding to the
vertex operator (\ref{eq:LCNS}) in the NS sector by
\begin{eqnarray}
V_{r\,\mathrm{L}}^{\left(-2\right)}(Z_{r}) & = & \frac{2}{p_{r}^{+}}ce^{-2\phi}\psi^{+}A_{-n_{1}}^{i_{1}(r)}\cdots B_{-s_{1}}^{j_{1}(r)}\cdots e^{-ip_{r}^{+}X_{\mathrm{L}}^{-}-i\left(p_{r}^{-}-\frac{2\mathcal{N}_{r}}{\alpha_{r}}-\frac{Q^{2}}{\alpha_{r}}\right)X_{\mathrm{L}}^{+}+ip_{r}^{i}X_{\mathrm{L}}^{i}}(Z_{r})~,\nonumber \\
V_{r\,\mathrm{L}}^{\left(-1\right)}(Z_{r}) & = & ce^{-\phi}A_{-n_{1}}^{i_{1}(r)}\cdots B_{-s_{1}}^{j_{1}(r)}\cdots e^{-ip_{r}^{+}X_{\mathrm{L}}^{-}-i\left(p_{r}^{-}-\frac{2\mathcal{N}_{r}}{\alpha_{r}}-\frac{Q^{2}}{\alpha_{r}}\right)X_{\mathrm{L}}^{+}+ip_{r}^{i}X_{\mathrm{L}}^{i}}(Z_{r})~,\nonumber \\
V_{r\,\mathrm{L}}^{\left(0\right)}(Z_{r}) & = & \left[cG_{-\frac{1}{2}}-\frac{1}{4}\gamma\right]A_{-n_{1}}^{i_{1}(r)}\cdots B_{-s_{1}}^{j_{1}(r)}\cdots e^{-ip_{r}^{+}X_{\mathrm{L}}^{-}-i\left(p_{r}^{-}-\frac{2\mathcal{N}_{r}}{\alpha_{r}}-\frac{Q^{2}}{\alpha_{r}}\right)X_{\mathrm{L}}^{+}+ip_{r}^{i}X_{\mathrm{L}}^{i}}(Z_{r})~,\label{eq:confNS}
\end{eqnarray}
with the DDF operators $A_{-n}^{i(r)},B_{-s}^{j(r)}$ for the $r$-th
string defined as
\begin{eqnarray}
A_{-n}^{i(r)} & = & \oint_{Z_{r}}\frac{d\mathbf{z}}{2\pi i}iD\left(\tilde{\mathcal{X}}^{i}+iQ\delta^{i,1}\Phi\right)e^{-i\frac{n}{p_{r}^{+}}\mathcal{X}_{\mathrm{L}}^{+}}(\mathbf{z})~,\nonumber \\
B_{-s}^{i(r)} & = & \oint_{Z_{r}}\frac{d\mathbf{z}}{2\pi i}\frac{D\mathcal{X}^{+}}{\left(ip_{r}^{+}\partial\mathcal{X}^{+}\right)^{\frac{1}{2}}}D\left(\tilde{\mathcal{X}}^{i}+iQ\delta^{i,1}\Phi\right)e^{-i\frac{s}{p_{r}^{+}}\mathcal{X}_{\mathrm{L}}^{+}}(\mathbf{z})~.\label{eq:B-s}
\end{eqnarray}
Here $\Phi$ is defined in (\ref{eq:Phi}) and $\mathcal{X}_{\mathrm{L}}^{+}$
denotes the left moving part of $\mathcal{X}^{+}$. The vertex operators
in (\ref{eq:confNS}) satisfy
\begin{eqnarray*}
XV_{r\,\mathrm{L}}^{\left(-2\right)}\left(Z_{r}\right) & = & V_{r\,\mathrm{L}}^{\left(-1\right)}(Z_{r})\,,\\
XV_{r\,\mathrm{L}}^{\left(-1\right)}(Z_{r}) & = & V_{r\,\mathrm{L}}^{\left(0\right)}\left(Z_{r}\right)\,,\\
Q_{\mathrm{B}}V_{r\,\mathrm{L}}^{\left(-2\right)}\left(Z_{r}\right) & = & Q_{\mathrm{B}}V_{r\,\mathrm{L}}^{\left(-1\right)}\left(Z_{r}\right)=Q_{\mathrm{B}}V_{r\,\mathrm{L}}^{\left(0\right)}\left(Z_{r}\right)=0\,,
\end{eqnarray*}
where $X$ is the picture changing operator (\ref{eq:PCO}) and $Q_{\mathrm{B}}$
denotes the BRST charge. The matter supercurrent $T_{\mathrm{F}}\left(z\right)$
satisfies the OPE
\begin{eqnarray*}
 &  & T_{\mathrm{F}}\left(z\right)A_{-n_{1}}^{i_{1}(r)}\cdots B_{-s_{1}}^{j_{1}(r)}\cdots e^{-ip_{r}^{+}X_{\mathrm{L}}^{-}-i\left(p_{r}^{-}-\frac{2\mathcal{N}_{r}}{\alpha_{r}}-\frac{Q^{2}}{\alpha_{r}}\right)X_{\mathrm{L}}^{+}+ip_{r}^{i}X_{\mathrm{L}}^{i}}(Z_{r})\\
 &  & \quad\sim\frac{1}{z-Z_{r}}G_{-\frac{1}{2}}A_{-n_{1}}^{i_{1}(r)}\cdots B_{-s_{1}}^{j_{1}(r)}\cdots e^{-ip_{r}^{+}X_{\mathrm{L}}^{-}-i\left(p_{r}^{-}-\frac{2\mathcal{N}_{r}}{\alpha_{r}}-\frac{Q^{2}}{\alpha_{r}}\right)X_{\mathrm{L}}^{+}+ip_{r}^{i}X_{\mathrm{L}}^{i}}(Z_{r})\,,
\end{eqnarray*}
with
\begin{eqnarray*}
 &  & G_{-\frac{1}{2}}A_{-n_{1}}^{i_{1}(r)}\cdots B_{-s_{1}}^{j_{1}(r)}\cdots e^{-ip_{r}^{+}X_{\mathrm{L}}^{-}-i\left(p_{r}^{-}-\frac{2\mathcal{N}_{r}}{\alpha_{r}}-\frac{Q^{2}}{\alpha_{r}}\right)X_{\mathrm{L}}^{+}+ip_{r}^{i}X_{\mathrm{L}}^{i}}(Z_{r})\\
 &  & \quad=\frac{p_{r}^{+}}{2}\psi^{-}A_{-n_{1}}^{i_{1}(r)}\cdots B_{-s_{1}}^{j_{1}(r)}\cdots e^{-ip_{r}^{+}X_{\mathrm{L}}^{-}-i\left(p_{r}^{-}-\frac{2\mathcal{N}_{r}}{\alpha_{r}}-\frac{Q^{2}}{\alpha_{r}}\right)X_{\mathrm{L}}^{+}+ip_{r}^{i}X_{\mathrm{L}}^{i}}(Z_{r})+\cdots\,.
\end{eqnarray*}
The ellipses on the right hand side denote the terms which do not
involve $\psi^{-}$. $V_{r\,\mathrm{L}}^{\left(-2\right)}\left(Z_{r}\right)$,
$V_{r\,\mathrm{L}}^{\left(-1\right)}\left(Z_{r}\right)$ and $V_{r\,\mathrm{L}}^{\left(0\right)}\left(Z_{r}\right)$
are the BRST invariant vertex operators in the $-2,\,-1,\,0$ pictures
respectively. 

The conformal gauge vertex operators corresponding to (\ref{eq:LCR})
in the R sector are defined to be
\begin{eqnarray}
V_{r\,\mathrm{L}}^{\left(-\frac{3}{2}\right)}(Z_{r}) & = & ce^{-\frac{3}{2}\phi}e^{\frac{i}{2}H}A_{-n_{1}}^{i_{1}(r)}\cdots B_{-s_{1}}^{j_{1}(r)}\cdots S_{\alpha}e^{-ip_{r}^{+}X_{\mathrm{L}}^{-}-i\left(p_{r}^{-}-\frac{2\mathcal{N}_{r}}{\alpha_{r}}-\frac{Q^{2}}{\alpha_{r}}\right)X_{\mathrm{L}}^{+}+ip_{r}^{i}X_{\mathrm{L}}^{i}}(Z_{r})~,\nonumber \\
V_{r\,\mathrm{L}}^{\left(-\frac{1}{2}\right)}(Z_{r}) & = & ce^{-\frac{1}{2}\phi}G_{0}e^{\frac{i}{2}H}A_{-n_{1}}^{i_{1}(r)}\cdots B_{-s_{1}}^{j_{1}(r)}\cdots S_{\alpha}e^{-ip_{r}^{+}X_{\mathrm{L}}^{-}-i\left(p_{r}^{-}-\frac{2\mathcal{N}_{r}}{\alpha_{r}}-\frac{Q^{2}}{\alpha_{r}}\right)X_{\mathrm{L}}^{+}+ip_{r}^{i}X_{\mathrm{L}}^{i}}(Z_{r})~.\label{eq:confR}
\end{eqnarray}
These are the vertex operators in the $-\frac{3}{2},\,-\frac{1}{2}$
pictures respectively and satisfy 
\begin{eqnarray*}
XV_{r\,\mathrm{L}}^{\left(-\frac{3}{2}\right)}(Z_{r}) & = & V_{r\,\mathrm{L}}^{\left(-\frac{1}{2}\right)}\left(Z_{r}\right)\,,\\
Q_{\mathrm{B}}V_{r\,\mathrm{L}}^{\left(-\frac{3}{2}\right)}\left(Z_{r}\right) & = & Q_{\mathrm{B}}V_{r\,\mathrm{L}}^{\left(-\frac{1}{2}\right)}\left(Z_{r}\right)=0\,.
\end{eqnarray*}
The matter supercurrent $T_{\mathrm{F}}\left(z\right)$ satisfies
the OPE
\begin{eqnarray*}
 &  & T_{\mathrm{F}}\left(z\right)e^{\frac{i}{2}H}A_{-n_{1}}^{i_{1}(r)}\cdots B_{-s_{1}}^{j_{1}(r)}\cdots S_{\alpha}e^{-ip_{r}^{+}X_{\mathrm{L}}^{-}-i\left(p_{r}^{-}-\frac{2\mathcal{N}_{r}}{\alpha_{r}}-\frac{Q^{2}}{\alpha_{r}}\right)X_{\mathrm{L}}^{+}+ip_{r}^{i}X_{\mathrm{L}}^{i}}(Z_{r})\\
 &  & \quad\sim\left(z-Z_{r}\right)^{-\frac{3}{2}}G_{0}e^{\frac{i}{2}H}A_{-n_{1}}^{i_{1}(r)}\cdots B_{-s_{1}}^{j_{1}(r)}\cdots S_{\alpha}e^{-ip_{r}^{+}X_{\mathrm{L}}^{-}-i\left(p_{r}^{-}-\frac{2\mathcal{N}_{r}}{\alpha_{r}}-\frac{Q^{2}}{\alpha_{r}}\right)X_{\mathrm{L}}^{+}+ip_{r}^{i}X_{\mathrm{L}}^{i}}(Z_{r})\,,
\end{eqnarray*}
with
\begin{eqnarray*}
 &  & G_{0}e^{\frac{i}{2}H}A_{-n_{1}}^{i_{1}(r)}\cdots B_{-s_{1}}^{j_{1}(r)}\cdots S_{\alpha}e^{-ip_{r}^{+}X_{\mathrm{L}}^{-}-i\left(p_{r}^{-}-\frac{2\mathcal{N}_{r}}{\alpha_{r}}-\frac{Q^{2}}{\alpha_{r}}\right)X_{\mathrm{L}}^{+}+ip_{r}^{i}X_{\mathrm{L}}^{i}}(Z_{r})\\
 &  & \quad=-\frac{p_{r}^{+}}{2}e^{-\frac{i}{2}H}A_{-n_{1}}^{i_{1}(r)}\cdots B_{-s_{1}}^{j_{1}(r)}\cdots S_{\alpha}e^{-ip_{r}^{+}X_{\mathrm{L}}^{-}-i\left(p_{r}^{-}-\frac{2\mathcal{N}_{r}}{\alpha_{r}}-\frac{Q^{2}}{\alpha_{r}}\right)X_{\mathrm{L}}^{+}+ip_{r}^{i}X_{\mathrm{L}}^{i}}(Z_{r})+\cdots\,.
\end{eqnarray*}
The ellipses on the right hand side denote the terms proportional
to $e^{\frac{i}{2}H}$. 

The right moving version of (\ref{eq:confNS}) and (\ref{eq:confR})
can be defined in the same way and the vertex operators in the $\left(p_{\mathrm{L}},\,p_{\mathrm{R}}\right)$
picture for Type II superstring theory are given in the form
\[
V_{r}^{\left(p_{\mathrm{L}},\,p_{\mathrm{R}}\right)}\left(Z_{r},\bar{Z}_{r}\right)=V_{r\,\mathrm{L}}^{\left(p_{\mathrm{L}}\right)}(Z_{r})V_{r\,\mathrm{R}}^{\left(p_{\mathrm{R}}\right)}(\bar{Z}_{r})\,.
\]

\section{A proof of the equality of (\ref{eq:FNgconftypeII}) and (\ref{eq:FNgconftypeII2})\label{subsec:A-proof-of}}

In this appendix, we show that (\ref{eq:FNgconftypeII}) is equal
to (\ref{eq:FNgconftypeII2}). This can be done by using fermionic
charges
\begin{eqnarray}
\hat{Q} & \equiv & \oint\frac{dz}{2\pi i}\left[-\frac{b}{4\partial\rho}\left(iX_{\mathrm{L}}^{+}-\frac{1}{2}\rho\right)\left(z\right)+\frac{\beta}{2\partial\rho}\psi^{+}\left(z\right)\right]\,,\nonumber \\
\hat{\bar{Q}} & \equiv & \oint\frac{d\bar{z}}{2\pi i}\left[-\frac{\bar{b}}{4\bar{\partial}\bar{\rho}}\left(iX_{\mathrm{R}}^{+}-\frac{1}{2}\bar{\rho}\right)\left(\bar{z}\right)+\frac{\bar{\beta}}{2\bar{\partial}\bar{\rho}}\bar{\psi}^{+}\left(\bar{z}\right)\right]\,,\label{eq:Qhat}
\end{eqnarray}
where
\begin{eqnarray}
\left(iX_{\mathrm{L}}^{+}-\frac{1}{2}\rho\right)\left(z\right) & \equiv & \int_{w_{0}}^{z}dz^{\prime}\left(i\partial X^{+}-\frac{1}{2}\partial\rho\right)\left(z^{\prime}\right)\,,\nonumber \\
\left(iX_{\mathrm{R}}^{+}-\frac{1}{2}\bar{\rho}\right)\left(\bar{z}\right) & \equiv & \int_{\bar{w}_{0}}^{\bar{z}}d\bar{z}^{\prime}\left(i\bar{\partial}X^{+}-\frac{1}{2}\bar{\partial}\bar{\rho}\right)\left(\bar{z}^{\prime}\right)\,,\label{eq:XLR}
\end{eqnarray}
with a generic point $w_{0}$ on the surface and the integration contours
on the right hand sides are taken to be the same. Since $\left(iX_{\mathrm{L}}^{+}-\frac{1}{2}\rho\right)\left(z\right),\,\left(iX_{\mathrm{R}}^{+}-\frac{1}{2}\bar{\rho}\right)\left(\bar{z}\right)$
thus defined depend on the contours and not single valued with respect
to $z,\bar{z}$, the right hand sides of (\ref{eq:Qhat}) are actually
ambiguous\footnote{The statement that $iX_{\mathrm{L}}^{+}-\frac{1}{2}\rho$ is single
valued made in appendix C.2 of \cite{Ishibashi2018} is wrong. However,
by replacing $\hat{Q}^{\prime}$ there by $\hat{Q}^{\prime}-\hat{\bar{Q}}^{\prime}$
which is well-defined, all the results there still hold as we show
in the following. }. Here let us consider the combination $\hat{Q}-\hat{\bar{Q}}$. One
can see that the ambiguity of this combination is proportional to
\[
\oint\frac{dz}{2\pi i}\frac{b}{\partial\rho}+\oint\frac{d\bar{z}}{2\pi i}\frac{\bar{b}}{\bar{\partial}\bar{\rho}}\,,
\]
which coincides with the $b_{0}-\bar{b}_{0}$ in light-cone gauge
string field theory. All the external states are annihilated by this
combination of antighosts and it is inserted in all the nontrivial
cycles. Therefore the combination $\hat{Q}-\hat{\bar{Q}}$ is well-defined
in the correlation functions discussed in this paper.

In order to use $\hat{Q}-\hat{\bar{Q}}$, we need to rewrite the ghost
part of the correlation function. Inserting
\[
1=\left|\oint_{w_{0}}\frac{dz}{2\pi i}\frac{b}{\partial\rho}\left(z\right)\partial\rho c\left(w_{0}\right)\right|^{2}\,,
\]
into (\ref{eq:FNgLCtypeII}) and deforming the contours of the antighost
insertions, (\ref{eq:FNgLCtypeII}) is transformed into 
\begin{eqnarray}
 &  & \int D\left[XBC\right]e^{-S^{\mathrm{tot}}}\partial\rho c\left(w_{0}\right)\bar{\partial}\bar{\rho}\bar{c}\left(\bar{w}_{0}\right)\nonumber \\
 &  & \qquad\times\prod_{J=1}^{g}\left[\left(\oint_{A_{J}}\frac{dz}{\partial\rho}b_{zz}+\oint_{\bar{A}_{J}}\frac{d\bar{z}}{\bar{\partial}\bar{\rho}}b_{\bar{z}\bar{z}}\right)\left(\oint_{B_{J}}\frac{dz}{\partial\rho}b_{zz}+\oint_{\bar{B}_{J}}\frac{d\bar{z}}{\bar{\partial}\bar{\rho}}b_{\bar{z}\bar{z}}\right)\right]\nonumber \\
 &  & \qquad\times\prod_{I}\left[\oint_{z_{I}}\frac{dz}{2\pi i}\frac{b}{\partial\rho}\left(z\right)X\left(z_{I}\right)\oint_{\bar{z}_{I}}\frac{d\bar{z}}{2\pi i}\frac{\bar{b}}{\partial\bar{\rho}}\left(\bar{z}\right)\bar{X}\left(\bar{z}_{I}\right)\right]\nonumber \\
 &  & \qquad\times\prod_{r}e^{-\frac{iQ^{2}}{\alpha_{r}}\mathcal{X}^{+}}\left(\hat{\tilde{\mathbf{z}}}_{I^{\left(r\right)}},\hat{\bar{\mathbf{z}}}_{I^{\left(r\right)}}\right)\prod_{r=1}^{N}V_{r}(Z_{r},\bar{Z}_{r})\,.\label{eq:FNgconftypeII3}
\end{eqnarray}
 The operators inserted at $z=z_{I}$ can be expressed as 
\begin{eqnarray}
 &  & \oint_{z_{I}}\frac{dz}{2\pi i}\frac{b}{\partial\rho}\left(z\right)X\left(z_{I}\right)\nonumber \\
 &  & \quad=-\oint_{z_{I}}\frac{dz}{2\pi i}\frac{b}{\partial\rho}\left(z\right)e^{\phi}T_{\mathrm{F}}^{\mathrm{LC}}\left(z_{I}\right)\nonumber \\
 &  & \hphantom{\quad=}-\left\{ \hat{Q}-\hat{\bar{Q}},\oint_{z_{I}}\frac{dz}{2\pi i}\frac{b}{\partial\rho}\left(z\right)\oint_{z_{I}}\frac{dw}{2\pi i}\frac{A\left(w\right)}{w-z_{I}}e^{\phi}\left(z_{I}\right)\right\} \nonumber \\
 &  & \hphantom{\quad=}+\frac{1}{4}\oint_{z_{I}}\frac{dz}{2\pi i}\frac{b}{\partial\rho}\left(z\right)\oint_{z_{I}}\frac{dw}{2\pi i}\frac{\partial\rho\psi^{-}\left(w\right)}{w-z_{I}}e^{\phi}\left(z_{I}\right)\,,\label{eq:bX}
\end{eqnarray}
where
\begin{eqnarray*}
A\left(w\right) & = & -i\partial X^{+}\partial\rho\gamma\left(w\right)-2\partial\left(\partial\rho c\right)\psi^{-}\left(w\right)\\
 &  & -\frac{d-10}{4}i\left[\left(\frac{5\left(\partial^{2}X^{+}\right)^{2}}{4\left(\partial X^{+}\right)^{3}}-\frac{\partial^{3}X^{+}}{2\left(\partial X^{;}\right)^{2}}\right)\left(-2\partial\rho\gamma\right)\right.\\
 &  & \hphantom{-\frac{d-10}{4}i\quad}\left.-\frac{2\partial^{2}X^{+}}{\left(\partial X^{+}\right)^{2}}\partial\left(-2\partial\rho\gamma\right)+\frac{\partial^{2}\left(-2\partial\rho\gamma\right)}{\partial X^{+}}-\frac{\left(-2\partial\rho\gamma\right)\partial\psi^{+}\partial^{2}\psi^{+}}{2\left(\partial X^{+}\right)^{3}}\right]\left(w\right)\,.
\end{eqnarray*}
Substituting (\ref{eq:bX}) into (\ref{eq:FNgconftypeII3}) and using
the fact that $\hat{Q}-\hat{\bar{Q}}$ commutes with other insertions,
we can show that (\ref{eq:FNgconftypeII3}) is equal to 
\begin{eqnarray}
 &  & \int D\left[XBC\right]e^{-S^{\mathrm{tot}}}\partial\rho c\left(w_{0}\right)\bar{\partial}\bar{\rho}\bar{c}\left(\bar{w}_{0}\right)\nonumber \\
 &  & \qquad\times\prod_{J=1}^{g}\left[\left(\oint_{A_{J}}\frac{dz}{\partial\rho}b_{zz}+\oint_{\bar{A}_{J}}\frac{d\bar{z}}{\bar{\partial}\bar{\rho}}b_{\bar{z}\bar{z}}\right)\left(\oint_{B_{J}}\frac{dz}{\partial\rho}b_{zz}+\oint_{\bar{B}_{J}}\frac{d\bar{z}}{\bar{\partial}\bar{\rho}}b_{\bar{z}\bar{z}}\right)\right]\nonumber \\
 &  & \qquad\times\prod_{I}\oint_{z_{I}}\frac{dz}{2\pi i}\frac{b}{\partial\rho}\left(z\right)\left[-e^{\phi}T_{\mathrm{F}}^{\mathrm{LC}}\left(z_{I}\right)+\frac{1}{4}\oint_{z_{I}}\frac{dw}{2\pi i}\frac{\partial\rho\psi^{-}\left(w\right)}{w-z_{I}}e^{\phi}\left(z_{I}\right)\right]\nonumber \\
 &  & \qquad\times\prod_{I}\oint_{\bar{z}_{I}}\frac{d\bar{z}}{2\pi i}\frac{\bar{b}}{\partial\bar{\rho}}\left(\bar{z}\right)\bar{X}\left(\bar{z}_{I}\right)\nonumber \\
 &  & \qquad\times\prod_{r}e^{-\frac{iQ^{2}}{\alpha_{r}}\mathcal{X}^{+}}\left(\hat{\tilde{\mathbf{z}}}_{I^{\left(r\right)}},\hat{\bar{\mathbf{z}}}_{I^{\left(r\right)}}\right)\prod_{r=1}^{N}V_{r}(Z_{r},\bar{Z}_{r})\,.\label{eq:FNgconfheterotic4}
\end{eqnarray}
It is easy to show that the term $\oint_{z_{I}}\frac{dw}{2\pi i}\frac{\partial\rho\psi^{-}\left(w\right)}{w-z_{I}}e^{\phi}\left(z_{I}\right)$
does not contribute to the correlation function \cite{Ishibashi2018}.
We can do the same thing for the antiholomorphic part and eventually
prove that (\ref{eq:FNgconftypeII}) is equal to (\ref{eq:FNgconftypeII2}).

Putting $Q=0$, we also find that (\ref{eq:confTypeII}) is equal
to (\ref{eq:conftypeII2}). 

\bibliographystyle{utphys}
\bibliography{SFT_18}

\end{document}